\documentclass[reprint, nofootinbib, aps,physrev]{revtex4-2}
\usepackage[english]{babel} 
\usepackage[T1]{fontenc}
\usepackage[utf8]{inputenc}
\usepackage{amsfonts,amsmath,amssymb,mathbbol,dsfont}
\usepackage{graphicx,psfrag,xcolor}
\usepackage[normalem]{ulem}
\usepackage{dcolumn}
\usepackage{bm} 
\usepackage{hyperref}
\usepackage{physics}
\definecolor{linkcolor}{HTML}{0176ba}
\definecolor{urlcolor}{HTML}{0176ba} 
\definecolor{citecolor}{HTML}{900020}
\hypersetup{pdfstartview=FitH,  linkcolor=linkcolor,urlcolor=urlcolor, citecolor=citecolor, colorlinks=true}
\usepackage[export]{adjustbox}
\usepackage{multirow}
\usepackage{array}
\usepackage{tikz}

\usepackage{listings}
\usepackage{color} 
\definecolor{mygreen}{RGB}{28,172,0} 
\definecolor{mylilas}{RGB}{170,55,241}
\definecolor{burntorange}{RGB}{204,85,0}
\usepackage{bbold}
\usepackage{mathtools}

\renewcommand{\Re}{\operatorname{Re}}
\renewcommand{\Im}{\operatorname{Im}}

\usepackage{comment} 
\excludecomment{hide}
\includecomment{keep}

\begin{document}

\title{Lattice-induced sound trapping in biperiodic metasurfaces of acoustic resonators}

\author{Nikita Ustimenko$^{1}$}
\email{nikita.ustimenko@kit.edu}
\author{Andrey B. Evlyukhin$^{2,3}$}
\author{Vicky Kyrimi$^4$}
\author{Alexander V. Kildishev$^5$}
 \author{Carsten Rockstuhl$^{1,6}$}
 \email{carsten.rockstuhl@kit.edu}
\affiliation{$^1$Institute of Theoretical Solid State Physics, Karlsruhe Institute of Technology, Kaiserstrasse 12, 76131 Karlsruhe, Germany}
\affiliation{$^2$Institute of Quantum Optics, Leibniz University Hannover, Welfengarten 1, 30167 Hannover, Germany}
\affiliation{{$^3$}Cluster of Excellence PhoenixD, Leibniz University Hannover, Welfengarten 1A, 30167 Hannover, Germany}
\affiliation{$^4$Department of Physics, Section of Condensed Matter Physics, National and Kapodistrian University of Athens, Panepistimioupolis, 157 84 Athens, Greece}
\affiliation{$^5$Elmore Family School of Electrical and Computer Engineering and Birck Nanotechnology Center, Purdue University, 1205 Mitch Daniels Blvd., West Lafayette, Indiana 47907-2057, USA}
\affiliation{$^6$Institute of Nanotechnology, Karlsruhe Institute of Technology, Kaiserstrasse 12, 76131 Karlsruhe, Germany}

\begin{abstract}
A referential example of a physical system that supports bound states in the continuum (BICs) with an infinite quality factor ($Q$ factor) is a metasurface of discrete scatterers (resonators), whose response can be significantly modified by exploiting lattice interactions. In this work, we explore the multipole-interference mechanism for realizing accidental acoustic BICs (trapped modes) at $\Gamma$-point (in-plane Bloch wave vector $\mathbf{k}_{\parallel} = \mathbf{0}$) of biperiodic metasurfaces of acoustic resonators with one resonator per unit cell. To do so, we expand the pressure field from the metasurface into a series of scalar zonal ($m = 0$) spherical multipoles, carried by a normally incident plane wave, and formulate analytical conditions on the resonator multipole moments under which an eigenmode becomes a BIC. The conditions enable us to determine the lattice constant and frequency values that facilitate the formation of an axisymmetric BIC with a specific parity, resulting from destructive interference between zonal multipoles of the same parity, despite each moment radiating individually. By employing the T-matrix method for acoustic metasurfaces, we numerically investigate the BIC resonance in various structures, including finite arrays, and also the transformation of such resonances into high-$Q$ quasi-BIC regimes, which can be excited by a plane wave at normal incidence.
\end{abstract}
 \maketitle

 \section{Introduction}

The physics of discrete electromagnetic or acoustic scatterers arranged in periodic lattices attracts significant attention because their electromagnetic or acoustic spectra can support additional collective resonances inaccessible with single scatterers~\cite{Utyushev2021Jun,Koshelev2021Jan,Babicheva2024Sep,Schulz2024Jun,Ma2014Sep}. Interactions between an infinite number of scatterers within a lattice can lead to the formation of lattice eigenmodes and significantly modify the lattice response compared to that of isolated scatterers or finite ensembles~\cite{Evlyukhin2010Jul,Luk'yanchuk2010Sep,allayarov2024multiresonant}. Among these eigenmodes, trapped modes or bound states in the continuum (BICs) are of particular interest in photonics and acoustics due to their unique characteristics~\cite{Evans1994Feb,Hsu2016Jul,Sadreev2021Apr,Koshelev2023May,Huang2024Jan,Huang2023Apr}. A genuine BIC has a frequency embedded in the continuum of radiating modes, but an infinite quality factor and no radiative losses~\cite{Hsu2013Jul}. This leads to confinement (trapping) of the mode energy near or inside the system~\cite{Huang2020Aug}.

The formation of BICs has been demonstrated in photonic and acoustic systems using various mechanisms~\cite{Marinica2008May,Hein2008Jun,Lyapina2015Oct,Koshelev2019Jun,Huang2021Aug,Deriy2022Feb,Krasikova2024Aug,He2025Jul}. One of these mechanisms involves coupling between discrete scatterers (particles) arranged in subwavelength biperiodic lattices, also referred to as \textit{metasurfaces}. The electromagnetic (in photonic or microwave systems) or acoustic (in acoustic systems) lattice-induced coupling can lead to the formation of a nonradiant eigenmode if the lattice constant is smaller than the wavelength in the background. In such a situation, a lattice eigenmode can radiate only in directions determined by its in-plane Bloch wave vector $\mathbf{k}_{\parallel}$. If the far field of the mode vanishes in these directions, the mode is commonly referred to as a symmetry-protected (or $\Gamma$-point) BIC for $\mathbf{k}_{\parallel} = \bm{0}$, and as an accidental (or parametric) BIC for $\mathbf{k}_{\parallel} \neq \bm{0}$~\cite{Sadrieva2019Sep}. However, this distinction is not exclusive, and accidental BICs have also been shown to occur in photonic systems at $\mathbf{k}_{\parallel} = \bm{0}$~\cite{Bulgakov2017Dec,Xiao2022May,Kostyukov2022Feb,allayarov2024anapole}. One approach involves forming a nonradiant state through destructive interference of multipoles with the same parity (symmetry with respect to inversion $\mathbf{r} \to -\mathbf{r}$)~\cite{allayarov2024anapole}.

A comprehensive description of BICs in metasurfaces can be obtained using the multipole decomposition method~\cite{Sadrieva2019Sep}, which has been successfully applied to describe lattice resonant effects in photonics and acoustics~\cite{allayarov2024anapole,Babicheva2021Jan,Liu2020May,Rahimzadegan2022May,Sainidou2005Mar,Li2024Aug}. This method expands an arbitrary field into mutually orthogonal functions with unique radiation patterns, called multipoles, providing a set of basic solutions to the Helmholtz equation~\cite{Bohren1998Apr}. Here, multipoles are assumed to be spherical waves. It is important to note that the multipole decomposition differs between electromagnetics and acoustics: electromagnetic waves are vectorial and transverse, whereas acoustic pressure waves are scalar and longitudinal. In electromagnetics, the decomposition involves vector electric and magnetic multipoles of degree $\ell \geq 1$, while in acoustics, it only involves scalar multipoles of degree $\ell \geq 0$. Thus, photonic systems possess two types of multipoles and resonances, providing multiple degrees of freedom to tailor the optical response~\cite{Fu2013Feb,Staude2013Sep,Babicheva2017Nov,Babicheva2018May}. Although acoustic scatterers possess only scalar multipoles, the presence of the monopole ($\ell = 0$), which can interfere with the dipole ($\ell = 1$) and higher-degree multipoles, gives rise to acoustics-specific effects, such as the acoustic Kerker effect~\cite{Wei2020Aug,Long2020Jun} and the lateral recoil force~\cite{Smagin2024Dec}.

The distinction outlined above necessitates different approaches to realize a BIC response in photonic and acoustic systems. The multipolar content of a symmetry-protected BIC (s-BIC) can contain only multipoles with $m \neq \pm 1$, which do not radiate in the normal direction with respect to the lattice~\cite{Sadrieva2019Sep}. The simplest example of an electromagnetic s-BIC is a lattice of vertically oriented electric or magnetic dipoles ($\ell = 1, m = 0$)~\cite{Sadrieva2019Sep,Koshelev2023May}. Such a dipole does not radiate transverse waves in the normal direction, making the entire lattice nonradiant if the dipole moments of all particles have the same amplitude and phase, \textit{i.e.}, $\mathbf{k}_{\parallel} = \bm{0}$. Moreover, such a mode can be excited under normal incidence only by breaking the particle symmetry~\cite{Fedotov2007Oct,Tuz2018Feb,Khardikov2012Feb,Koshelev2018Nov,Barreda2022Aug}, which breaks inversion and rotational symmetry with respect to the $z$ axis~\cite{Poleva2023Jan}. This can be achieved, for example, by replacing a sphere with a non-equilateral prism. This symmetry breaking enables bianisotropy-induced coupling between the nonradiant electric dipole from the s-BIC with $m = 0$ and the magnetic dipole of the external wave with $m = \pm 1$, or vice versa~\cite{Evlyukhin2020May,Evlyukhin2021Dec}. This state is referred to as a quasi-s-BIC to emphasize its finite but large $Q$ factor. A quasi-BIC can be more practically important than a genuine BIC, as it can be excited by an external plane wave at normal incidence~\cite{Koshelev2019Jul,Carletti2018Jul}.

In acoustics, a normally incident pressure plane wave is a superposition of scalar multipoles that always radiate in the normal direction -- so-called \textit{zonal multipoles} with $m = 0$. Although scalar multipoles with $m \neq 0$ do not radiate in the normal direction and can form acoustic s-BICs, it can be hindered, if not impossible, to efficiently excite an acoustic quasi-s-BIC with a high $Q$-factor under normal incidence without breaking the particle rotational symmetry~\cite{Tsimokha2022Apr}. Recently, Allayarov \textit{et al}. showed that lattice-induced coupling between multipoles of the same parity (and $m = \pm 1$) can enable accidental optical BICs (a-BICs) at the $\Gamma$-point of a metasurface of spherical resonators~\cite{allayarov2024anapole}. In this case, each multipole can be excited by an external wave, resulting in quasi-a-BICs with high-$Q$ resonances under normal incidence without breaking metasurface symmetry.

\begin{figure}
     \centering
     \includegraphics[scale=0.29]{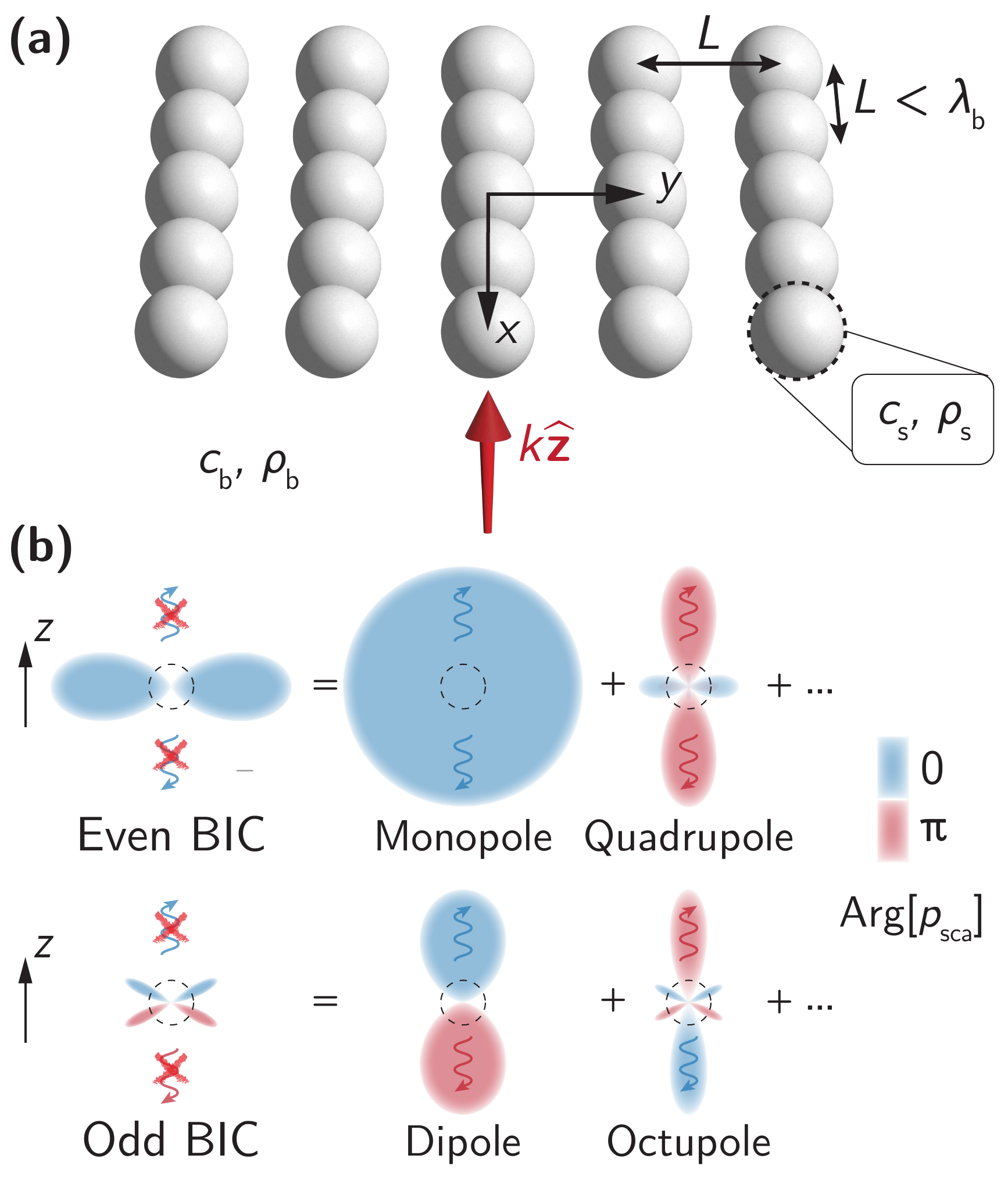}
     \caption{(a) Sketch of a biperiodic lattice of spherical particles (resonators) with a subwavelength lattice constant of $L$; the structure supports accidental acoustic bound states in the continuum (trapped modes) with a certain parity at the $\Gamma$ point (normal incidence).  (b) Formation of the acoustic BICs owing to destructive interference between scalar zonal multipoles with the same parity (with respect to $\mathbf{r} \to -\mathbf{r}$). The diagrams illustrate the intensity of the pressure field in the far-field region, $r \gg \lambda_{\rm b}$, generated by a unit cell. The color represents the pressure field phase, while the radius indicates its amplitude.}
     \label{fig:sketch}
 \end{figure}

Therefore, in this work, we extend and explore this approach for acoustic metasurfaces excited by a normally incident pressure wave [see Fig.~\ref{fig:sketch}(a)], showing its consistency when mapped onto a different wave propagation platform. We demonstrate that metasurfaces with inversion symmetry can also support accidental acoustic BICs and quasi-BICs for $\mathbf{k}_{\parallel} = \bm{0}$, arising from monopole–quadrupole or dipole–octupole coupling that suppresses acoustic radiation in the normal direction [see Fig.~\ref{fig:sketch}(b)]. In Sec.~\ref{sec:model}, we present a comprehensive theoretical model to describe the acoustic response of metasurfaces and their eigenmodes. This model enables us to derive in Sec.~\ref{sec:bic}, analytical conditions under which axisymmetric acoustic BICs emerge. Moreover, in Sec.~\ref{sec:bic}, we model them in infinite arrays of spherical resonators and discuss their transformation into quasi-BICs as the lattice constant and frequency are tuned. Finally, in Sec.~\ref{sec:quasi-bic}, we investigate the quasi-BIC regime under realistic conditions, such as the presence of material losses in the particles or a substrate, as well as finite array sizes.

 \section{Theoretical model}\label{sec:model}
\subsection{Basic equations}
In linear acoustics, the propagation of a monochromatic acoustic wave in a medium with compressibility $\beta$ and mass density $\rho$ is described by pressure $p(\omega; \mathbf{r})$ and velocity $\mathbf{v}(\omega;\mathbf{r})$ that obey the laws of mass and linear momentum conservation~\cite{Bruus2011Dec,Williams1999}:
\begin{align}
\label{eq_basic}
    i\omega \beta p(\omega; \mathbf{r}) = \bm{\nabla} \cdot \mathbf{v}(\omega;\mathbf{r}), \ i\omega \rho \mathbf{v}(\omega; \mathbf{r}) = \bm{\nabla} p(\omega;\mathbf{r})\, .
\end{align}
By substituting the second equation into the first one, we see that the pressure field obeys the acoustic wave equation,
\begin{align}
\label{wave_eq}
    \Delta p(\omega; \mathbf{r}) + \frac{\omega^2}{c^2}  p(\omega; \mathbf{r}) = 0\, ,
\end{align}
where $c = 1/\sqrt{\beta\rho}$ is the speed of sound. Formally, Eq.~(\ref{wave_eq}) presents the homogeneous Helmholtz equation with wavenumber $k = \omega/c$.

\subsection{T-matrix approach to scattering of acoustic waves}\label{sec:theory}
\subsubsection{Single resonator}
Let us consider the scattering of an acoustic wave $p_{\rm inc}(\omega;\mathbf{r})$ by an acoustic resonator. The linearity of Eq.~\eqref{wave_eq} allows us to write the total pressure field in the system as the sum of the incident and scattered pressure fields,
\begin{align}
\label{eq_ptot}
    p_{\rm tot}(\omega; \mathbf{r}) = p_{\rm inc}(\omega; \mathbf{r}) + p_{\rm sca}(\omega; \mathbf{r})\, .
\end{align}
An efficient approach to calculate the scattered field in spherical coordinates, which involves expanding $p_{\rm inc}(\omega; \mathbf{r})$ and $p_{\rm sca}(\omega; \mathbf{r})$ into a series of fundamental solutions of Eq.~\eqref{wave_eq}, was laid out by Lord Rayleigh in his seminal work, \textit{The Theory of Sound} (see Chap. 17 of Ref.~\onlinecite{Rayleigh1896}). In today`s notation, such fundamental solutions, the scalar spherical waves or simply multipoles $\Psi^{(n)}_{\ell, m}(\omega; \mathbf{r})$ with $\ell=0,1,2,...$ and $m=-\ell, -\ell+1,...,\ell-1,\ell$ being the degree and order~\cite{Anderson1950Jul}, are commonly used to model acoustic sources and scatterers. The definition of $\Psi^{(n)}_{\ell, m}(\omega; \mathbf{r})$ is given in Appendix~\ref{sec:ssw}. In the case of the acoustic scattering by an individual particle (resonator), the series takes the following form (we assume that the series converges to the actual fields outside the smallest sphere encapsulating the entire scatterer):
\begin{align}
    \label{eq_pinc}
    p_{\rm inc}(\omega; \mathbf{r}) &= \sum_{\ell,m}  b_{\ell, m}(\omega) \Psi^{(1)}_{\ell, m}(\omega; \mathbf{r} - \mathbf{r}_0)\, , \\
    \label{eq_psca}
     p_{\rm sca}(\omega; \mathbf{r}) &= \sum_{\ell,m}  a_{\ell, m}(\omega) \Psi^{(3)}_{\ell, m}(\omega; \mathbf{r} - \mathbf{r}_0)\, ,
\end{align}
where $\sum_{\ell,m} \equiv \sum_{\ell =0}^{+\infty} \sum_{m=-\ell}^{\ell}$. The arguments of the expansion coefficients include the origin of the multipole expansion, denoted here as $\mathbf{r}_0$. Following Ref.~\onlinecite{Evlyukhin2011Dec}, we assume that $\mathbf{r}_0$ coincides with the center of mass of the scatterer and the origin of the coordinate system to omit the dependence on $\mathbf{r}_0$ further. To highlight the dependence of the coefficients on the frequency $\omega$ in the general case, we explicitly write it as an argument. However, for a pressure plane wave, the coefficients $b_{\ell, m}(\omega)$ do not depend on the frequency $\omega$, but depend on the direction of incidence (see Appendix~\ref{sec:plane_wave_exp}). 

The expansion coefficients $a_{\ell, m}(\omega)$ of the scatterer can be calculated through the expansion coefficients $b_{\ell, m}(\omega)$ of the source and the transition matrix (T-matrix) of the scatterer as
\begin{align}
    a_{\ell, m}(\omega) = \sum_{\ell',m'} T_{\ell, m,\ell',m'}(\omega) b_{\ell',m'}(\omega)\, ,
\end{align}
or, in matrix form,
\begin{align}
\label{eq_Tmatrix}
    \mathbf{a}(\omega) = \mathbf{T}(\omega) \mathbf{b}(\omega)\, ,
\end{align}
where $\mathbf{a}(\omega)$ and $\mathbf{b}(\omega)$ are the column vectors of the corresponding coefficients arranged in ascending order of $\ell = 0,1,2,...$ and $m=-\ell,-\ell+1,...,\ell$, and $\mathbf{T}(\omega)$ is the T-matrix after a truncation of the series at $\ell_{\rm max}$ (large enough to reach convergence). The T-matrix contains the necessary and sufficient ``information'' about the scatterer to describe the scattering of an arbitrary field $p_{\rm inc}(\omega; \mathbf{r})$ at a given frequency~\cite{Waterman1969Jun,Waterman2009Jan}. The elements of the T-matrix are calculated analytically for the spherical scatterers (see Ref.~\onlinecite{Sainidou2005Mar} and also Appendix~\ref{sec:tm_sphere}) and numerically for arbitrarily shaped scatterers~\cite{Ustimenko2025Apr}.

\subsubsection{Biperiodic metasurface of resonators}
The T-matrix method also provides an analytical, clear, and efficient treatment of the scattering of a plane wave by periodic metasurfaces~\cite{Necada2021Jan,Rahimzadegan2022May}. We assume a lattice in the $xy$ plane with a simple unit cell, \textit{i.e.}, one acoustic scatterer per unit cell [see Fig.~\ref{fig:sketch}(a)]. First, we notice that, because of the translation symmetry, the multipole moments of the unit cell at $\mathbf{R}$ are linked to those of the reference unit cell at $\bm{0}$ via Bloch's theorem, 
\begin{align}
\label{eq_bloch}
    \mathbf{a}^{\rm eff}_{\mathbf{R}} =  \mathrm{e}^{\mathrm{i} \mathbf{k}_{\parallel}\cdot\mathbf{R}}  \mathbf{a}^{\rm eff}\, ,
\end{align}
where the in-plane wave vector $\mathbf{k}_{\parallel}$ determines the phase shift between the unit cells. In the presence of an incident plane wave, $\mathbf{k}_{\parallel}$ is determined by the in-plane component of its wave vector, \textit{i.e.} $\mathbf{k} = \mathbf{k}_{\parallel} \pm \sqrt{k^2-k_{\parallel}^2} \:\widehat{\mathbf{z}}$, where $\widehat{\mathbf{z}}$ is the unit out-of-plane vector directed along the $z$ axis of the Cartesian coordinate system in Fig.~\ref{fig:sketch}(a). In the case of an eigenmode, $\mathbf{k}_{\parallel}$ is its Bloch wave vector from the first Brillouin zone. A subscript ``eff'' is introduced to distinguish the multipole moments of the scatterer embedded in the lattice~\eqref{eq_a_eff} from the moments ${\bf a}(\omega)$ of the isolated scatterer from Eq.~(\ref{eq_Tmatrix}). This difference arises because, when a scatterer is placed on a lattice, the scatterers interact with each other. Hence, the unit cell with one scatterer placed at $\mathbf{r}_0 = \bm{0}$ (reference unit cell) experiences (i) the incident field $p_{\rm inc}(\omega;\mathbf{r})$, and (ii) the secondary fields, rescattered by all the other scatterers located at the lattice nodes $\mathbf{R} \neq \bm{0}$ and determined by the multipole coefficients $\mathbf{a}^{\rm eff}_{\mathbf{R}}$. The lattice-induced interaction between the unit cells can be accommodated within the T-matrix method by multiplication with the lattice sum matrix $\bm{\Sigma}'(\omega,\mathbf{k}_{\parallel})$ in Eq.~\eqref{eq_Tmatrix}, where $'$ denotes that the reference unit cell is excluded from the sum [see Eq.~\eqref{eq_lattice_sums}]. Hence, the multipole coefficients of the reference unit cell can be computed by solving the following equation (see Appendix~\ref{sec:lattice} for details)   
\begin{align}
\label{eq_mse_lattice}
    \mathbf{a}^{\rm eff} =  \mathbf{T}(\omega) \left[\mathbf{b} + \bm{\Sigma}'(\omega,\mathbf{k}_{\parallel})\mathbf{a}^{\rm eff}\right] 
\end{align}
or
\begin{align}
\label{eq_mse_lattice_2}
    \underbrace{\mathbf{T}^{-1}(\omega) \left[ \mathbf{I} - \mathbf{T}(\omega) \mathbf{\Sigma}'(\omega,\mathbf{k}_{\parallel}) \right]}_{\mathbf{T}^{-1}_{\rm eff}(\omega,\mathbf{k}_{\parallel})} \mathbf{a}^{\rm eff} = \mathbf{b}\, , 
\end{align}
where $\mathbf{I}$ is the identity matrix of the required size, $\mathbf{T}_{\rm eff}(\omega,\mathbf{k}_{\parallel}) \equiv \left[ \mathbf{I} - \mathbf{T}(\omega) \mathbf{\Sigma}'(\omega,\mathbf{k}_{\parallel}) \right]^{-1}\mathbf{T}(\omega)$ is the effective acoustic T-matrix of the resonator on the lattice, and $\mathbf{b}$ contains the frequency-independent expansion coefficients of a plane wave into scalar spherical multipoles (see Appendix~\ref{sec:plane_wave_exp}). 

Equation~\eqref{eq_mse_lattice_2} can be used to compute the acoustic response of a metasurface to an external wave (\textit{e.g.}, plane wave). In this case, a formal solution to Eq.~\eqref{eq_mse_lattice_2} is given by 
\begin{align}
\label{eq_a_eff}
    \mathbf{a}^{\rm eff} = \mathbf{T}_{\rm eff}(\omega,\mathbf{k}_{\parallel})\mathbf{b} =   \left[ \mathbf{I} - \mathbf{T}(\omega) \mathbf{\Sigma}'(\omega,\mathbf{k}_{\parallel}) \right]^{-1}\mathbf{a}\, ,
\end{align}
 where we have used Eq.~(\ref{eq_Tmatrix}). To recall the dependence of the vector of effective multipole coefficients on $\omega$ and $\mathbf{k}_{\parallel}$, we write $\mathbf{a}^{\rm eff}$ as $\mathbf{a}^{\rm eff} (\omega,\mathbf{k}_{\parallel})$ if such a highlight is needed. 
 
 Note that while vector $\mathbf{a}$ of the single scatterer in Eq.~\eqref{eq_a_eff} depends on $\omega$, it also depends on $\mathbf{k}_{\parallel}$, but only because the expansion of a plane wave into spherical multipoles $\mathbf{b}$ depends on $\mathbf{k}$ (see Appendix~\ref{sec:plane_wave_exp}). In contrast to the single-particle T-matrix $\mathbf{T}(\omega)$, which is calculated once at a given frequency and used for an arbitrary incident wave, the matrix $\mathbf{T}_{\rm eff}(\omega,\mathbf{k}_{\parallel})$ in Eq.~\eqref{eq_a_eff} must be recomputed for all the desired values of $\mathbf{k}_{\parallel}$ at a given frequency. 
 
 It also follows from Eq.~\eqref{eq_a_eff} that the coupling between an infinite number of scatterers within a metasurface leads to the renormalization of their acoustic multipole response $\mathbf{a}^{\rm eff} = \left[ \mathbf{I} - \mathbf{T}(\omega) \mathbf{\Sigma}'(\omega,\mathbf{k}_{\parallel}) \right]^{-1}\mathbf{a}$. The scattered (radiated) pressure field for a metasurface is a superposition of the fields emitted by an infinite number of particles on the lattice [see, \textit{e.g.}, Eq.~\eqref{eq_p_lattice}]. 
When the determinant of $\mathbf{T}_{\rm eff}^{-1}$ in Eq.~\eqref{eq_mse_lattice_2} approaches zero, we observe resonant features in the spectrum of the metasurface. A detailed discussion of these features, related to the lattice eigenmodes~\cite{Necada2021Jan,Li2024Aug,Beutel2024Apr}, is given in Sec.~\ref{sec:bic}.

\subsection{Selection rules for multipole coupling in metasurfaces}\label{subsec:rules}
Further consideration is facilitated by the symmetry analysis and associated selection rules that constrain which multipoles interact in structures with a given symmetry~\cite{Tsimokha2022Apr,Gladyshev2020Aug,Poleva2023Jan}, or in other words, which off-diagonal elements of $\mathbf{T}^{-1}_{\rm eff}(\omega,\mathbf{k}_{\parallel})$ in Eq.~\eqref{eq_mse_lattice_2} do not vanish.

Let us first analyze the acoustic coupling of scalar multipoles induced by a two-dimensional (2D) Bravais lattice with rotational symmetry of order $\mathfrak{n}_{\rm L}$, considering rotations by $2\pi/\mathfrak{n}_{\rm L}$ around the $z$ axis. We also assume that each unit cell contains only one sphere, for which the single-particle T-matrix is diagonal $T_{\ell,m,\ell',m'} = \mathfrak{a}_{\ell}\delta_{\ell,\ell'} \delta_{m,m'}$, where $\delta_{\ell,\ell'}$ is the Kronecker symbol (see Appendix~\ref{sec:tm_sphere} for the definition of $\mathfrak{a}_\ell$). Hence, the diagonal elements of $ \mathbf{T}^{-1}_{\rm eff}$ become $\left(T^{-1}_{\ell,m,\ell,m} - \Sigma'_{\ell,m,\ell,m}\right)$, while the off-diagonal elements are $\left(-\Sigma'_{\ell,m,\ell',m'}\right)$ with $\ell \neq \ell'$ and $m \neq m'$. Thus, the metasurface has inversion symmetry (considering $\mathbf{r} \to -\mathbf{r}$) and rotational symmetry of order $\mathfrak{n}_{\rm L}$. Lattice-induced coupling between two different multipoles $(\ell,m)$ and $(\ell',m')$ is allowed ($\Sigma'_{\ell,m,\ell',m'} \neq 0$) for an arbitrary $\mathbf{k}_{\parallel}$ if $\left(\ell+\ell'\right) = (m-m') \ \text{mod} \ 2$~\cite{Beutel2023Jan}. Under normal incidence $\mathbf{k}_{\parallel} = \bm{0}$, the following conditions must also be satisfied (see Appendix~\ref{sec:selection_rules}):
    \begin{align}
    \label{eq_selection_rules}
        \ell=\ell' \ \text{mod} \ 2\, , \quad \quad
        m = m' \ \text{mod} \ \mathfrak{n}_{\rm L}\, .
    \end{align}
The first condition states that the multipoles must have the same parity, which follows from the inversion symmetry of the metasurface, while the second condition follows from its rotational symmetry.

However, even though $\Sigma'_{\ell,m,\ell',m'} = 0$, the \textit{particle-induced} coupling due to off-diagonal entries of the T-matrix $T_{\ell,m,\ell',m'} \neq 0$ can be present for nonspherical acoustic resonators.  Shortly, the inversion symmetry of the particle requires $\ell=\ell' \ \text{mod} \ 2$, while the rotational symmetry of order $\mathfrak{n}_{\rm P}$ implies $m = m' \ \text{mod} \ \mathfrak{n}_{\rm P}$ [thus, the unit cell has the rotational symmetry of order $\mathfrak{n} = \min\left(\mathfrak{n}_{\rm L}, \mathfrak{n}_{\rm P}\right)$]. Note that resonators can have other symmetries (\textit{e.g.}, reflection), which determine additional selection rules but are not considered here.  The full list of selection rules for single acoustic resonators of different symmetries can be found in Ref.~\onlinecite{Tsimokha2022Apr}.

\subsection{Reflection and transmission coefficients}
Effective multipole moments~\eqref{eq_a_eff} allow us to calculate the reflection and transmission coefficients of a metasurface for an acoustic plane wave. To do this, we consider the pressure field in the far-field zone for $z < 0$ and $z > 0$. The incident wave is a pressure plane wave $p_{\rm inc}(\omega;\mathbf{r}) = p_0 \mathrm{e}^{\mathrm{i} \mathbf{k} \cdot \mathbf{r}}$ with $\mathbf{k} = \mathbf{k}_{\parallel} + \sqrt{k^2 - k_{\parallel} ^2}\hat{\mathbf{z}}$ and $|\mathbf{k}| =k$. Using Eqs.~\eqref{eq_ptot} and~\eqref{eq_psca} with multipole coefficients~\eqref{eq_a_eff}, one can write the total pressure after the summation over the lattice sites as
\begin{align}
\label{eq_p_lattice}
    p_{\rm tot}(\omega;\mathbf{r}) = p_0 \left[ \mathrm{e}^{\mathrm{i} \mathbf{k} \cdot \mathbf{r}} + \sum_{\ell,m}  a_{\ell,m}^{\rm eff} \Sigma_{\ell, m}\right]\, ,
\end{align}
where the lattice field sum is
\begin{align}
\label{eq_sigma}
   \Sigma_{\ell, m}(\omega, \mathbf{k}_{\parallel}; \mathbf{r}) = \sum_{\mathbf{R}} \Psi_{\ell,m}^{(3)}(\omega; \mathbf{r}- \mathbf{R})  \mathrm{e}^{\mathrm{i}\mathbf{k}_{\parallel} \cdot \mathbf{R}}\, .
\end{align}
Note that the second term in the brackets in Eq.~\eqref{eq_p_lattice} corresponds to the scattered field in the presence of an external field $p_0 \neq 0$ and the field of a lattice eigenmode in the absence $p_0 = 0$ (see Sec.~\ref{sec:bic_condition}). 

If the lattice constant is smaller than the wavelength in the background medium with $c_{\rm b}$, $L < \lambda_{\rm b}$, lattice field sums~\eqref{eq_sigma} can be written in the far-field region ($r \gg \lambda_{\rm b}$) as (see Appendix~\ref{sec:sums})
\begin{align}
\label{eq_sigma_lm_subwave}
    \Sigma_{\ell, m}(\omega, \mathbf{k}_{\parallel}; \mathbf{r}) = \frac{2 \pi}{k^2 A} (-\mathrm{i})^{\ell} Y_{\ell, m}(\theta_{\mathbf{k}_{\pm}},\varphi_{\mathbf{k}_{\pm}}) \mathrm{e}^{\mathrm{i}\mathbf{k}_{\pm}\cdot\mathbf{r}}\, ,
\end{align}
where the wave vectors are $\mathbf{k}_{\pm} = \mathbf{k}_{\parallel} \pm \sqrt{k^2 - k_{\parallel} ^2}\hat{\mathbf{z}}$ for $z \gtrless 0$ and $A$ is the area of a lattice unit cell ($A = L^2$ for a square unit cell). By inserting Eq.~\eqref{eq_sigma_lm_subwave} into Eq.~\eqref{eq_p_lattice}, we obtain the pressure field reflection and transmission coefficients for a metasurface as
\begin{subequations}
    \begin{align}
    \label{eq_r}
    r(\omega, \mathbf{k}_{\parallel}) &= \mathcal{P}_-(\omega, \mathbf{k}_{\parallel})\, , \\
    \label{eq_t}
    t(\omega, \mathbf{k}_{\parallel})  &= 1 + \mathcal{P}_+(\omega, \mathbf{k}_{\parallel})\, ,
\end{align}
\end{subequations}
while the intensity reflectance and transmittance are given by $R = |r|^2$ and $T = |t|^2$, respectively. Here, the far-field amplitudes of the scattered waves for $z \gtrless 0$ are
\begin{align}
\label{eq_P}
    \mathcal{P}_{\pm}(\omega, \mathbf{k}_{\parallel}) = \frac{2 \pi}{k^2 A}\sum_{\ell, m} (-\mathrm{i})^{\ell} a_{\ell,m}^{\rm eff} Y_{\ell, m}(\theta_{\mathbf{k}_{\pm}},\varphi_{\mathbf{k}_{\pm}})\, .
\end{align}
The quantities in Eqs.~\eqref{eq_r}-\eqref{eq_P} depend on $\omega$ and $\mathbf{k}_{\parallel}$ due to $a_{\ell,m}^{\rm eff} \equiv a_{\ell,m}^{\rm eff}(\omega, \mathbf{k}_{\parallel})$. In Eq.~\eqref{eq_P}, $\mathbf{k}_{+}$ points into the direction of propagation of the incident and transmitted plane waves, while $\mathbf{k}_{-}$ is the wave vector of the reflected wave. 
Thus, a subwavelength biperiodic lattice can radiate only in two directions $\mathbf{k}_{-}$ and $\mathbf{k}_{+}$ with the amplitude and phase dictated by $\mathcal{P}_-$ and $\mathcal{P}_+$, respectively. The response of the lattice is determined by the effective multipole moments, \textit{i.e.}, the renormalized response of one unit cell~\eqref{eq_a_eff}. Moreover, the lattice coupling does not change the angular dependence of the multipoles of a unit cell, given by $Y_{\ell, m}(\theta_{\mathbf{k}_{\pm}},\varphi_{\mathbf{k}_{\pm}})$, but only changes the multipole moments from Eqs.~\eqref{eq_Tmatrix} to~\eqref{eq_a_eff}. Thus, one can control the radiation pattern of the unit cell by tailoring the values of the multipole coefficients. Note also that, in the absence of the incident field $\mathbf{b} = \bm{0}$, expressions~\eqref{eq_P} describe the far-field amplitudes of the scattered waves radiated by a lattice eigenmode (see Sec.~\ref{sec:bic_condition}). 

Finally, let us consider the case of normal incidence $\mathbf{k}_{\parallel} = \bm{0}$. For the arguments $\cos \theta_{\pm} = \pm1$ corresponding to $\theta = 0$ ($+z$ direction) and $\theta = \pi$ ($-z$ direction), the associated Legendre polynomials~\eqref{eq_Plm} take the following values~\cite{Rahimzadegan2022May}:
\begin{align}
\label{eq_Plm_1}
    P^m_{\ell}(\pm1) & = 
    \begin{cases}
        0,& m \neq 0\,, \\
        (\pm 1)^{\ell},& m = 0\,.
    \end{cases}
\end{align}
From Eq.~\eqref{eq_P}, we see that for $\mathbf{k}_{\parallel} = \bm{0}$, the amplitudes of the scattered (radiated) fields separate into two subseries corresponding to even and odd multipoles, in full agreement with the first selection rule from Eq.~\eqref{eq_selection_rules},
\begin{align}
\begin{aligned}
\label{eq_P_zonal}
\mathcal{P}_{\pm} = \frac{\sqrt{\pi}}{k^2 A}\sum_{\text{even} \ \ell} (-\mathrm{i})^{\ell}\sqrt{2\ell + 1}  a_{\ell,0}^{\rm eff} \\
    \pm \frac{\sqrt{\pi}}{k^2 A} \sum_{\text{odd} \ \ell} (-\mathrm{i})^{\ell}\sqrt{2\ell + 1}  a_{\ell,0}^{\rm eff}\, ,
\end{aligned}
\end{align}
where $\mathcal{P}_{\pm} \equiv \mathcal{P}_{\pm}(\omega,\bm{0})$ and $a_{\ell,0}^{\rm eff} \equiv a_{\ell,0}^{\rm eff}(\omega,\bm{0})$ for compactness. Thus, the response of a metasurface at normal incidence can be considered as a superposition of the responses of two noninteracting sublattices, each of which includes either only even multipoles or only odd multipoles, respectively.

\section{Accidental bound states in the continuum}\label{sec:bic}

\subsection{Eigenmode and BIC condition}\label{sec:bic_condition}

Lattice eigenmodes can be obtained by solving Eq.~\eqref{eq_mse_lattice_2} in the absence of external fields $(\mathbf{b} = \bm{0})$,
\begin{align}
\label{eq_lattice_modes}
   \mathbf{T}^{-1}_{\rm eff}(\omega,\mathbf{k}_{\parallel})\mathbf{a}^{\rm eff} = \bm{0}\, .
\end{align}
According to Cramer`s theorem, nontrivial solutions to homogeneous Eq.~\eqref{eq_lattice_modes} exist if the inverse effective T-matrix is singular, \textit{i.e.}, $\det \mathbf{T}^{-1}_{\rm eff}(\omega,\mathbf{k}_{\parallel}) = 0$~\cite{Strang2023}. The singularity of the inverse effective T-matrix indicates that the system has reached a resonance mode where a self-sustaining solution can exist without external stimulus. 

For an open system and real $\mathbf{k}_{\parallel}$, a nontrivial solution to Eq.~\eqref{eq_lattice_modes} can generally be found for a complex frequency $[\omega(\mathbf{k}_{\parallel})-\mathrm{i}\gamma_{\rm r}(\mathbf{k}_{\parallel}) - \mathrm{i}\gamma_{\rm nr}(\mathbf{k}_{\parallel})]$, where $\gamma_{\rm r}(\mathbf{k}_{\parallel})$ and $\gamma_{\rm nr}(\mathbf{k}_{\parallel})$ define energy dissipation due to radiative and nonradiative losses, respectively. However, photonic and acoustic metasurfaces can support eigenmodes without radiative losses $(\gamma_{\rm r} = 0)$, which are just called \textit{bound states in the continuum or BICs}. Moreover, in the absence of nonradiative losses (absorption), BICs emerge for pure real-valued frequencies due to $\gamma_{\rm nr} = 0$. Hence, a criterion of the BIC occurrence can be formulated as~\cite{Bulgakov2017Dec,Bulgakov2018Oct} 
\begin{align}
\label{eq_det}
     \det \mathbf{T}^{-1}_{\rm eff}(\omega,\mathbf{k}_{\parallel}) = 0\, , \quad \quad \omega \in \mathbb{R}\, .
\end{align}

Since the matrix $\mathbf{T}^{-1}_{\rm eff}(\omega,\mathbf{k}_{\parallel})$ can be subject to the singular value decomposition (SVD),  the absolute value of its determinant is equal to the product of its singular values, which are nonnegative real-valued numbers~\cite{Suryadharma2017May}.
Thus, if a singular value of the matrix is zero, then the determinant is also zero. Hence, the solution to Eq.~\eqref{eq_lattice_modes} is the right singular vector of matrix $\mathbf{T}^{-1}_{\rm eff}(\omega,\mathbf{k}_{\parallel})$ with a singular value equal to zero~\cite{Necada2021Jan,Wang2025Feb}. After denoting the operator that picks the smallest singular value of a matrix with $s_{\rm min}$, BIC condition~\eqref{eq_det} reads as
\begin{align}
\label{eq_smin}
 s_{\rm min}\left[\mathbf{T}^{-1}_{\rm eff}(\omega,\mathbf{k}_{\parallel}) \right] = 0\, , \quad \quad \omega \in \mathbb{R}\, .
\end{align}
Conditions~\eqref{eq_det} and~\eqref{eq_smin} ensure the existence of a lattice eigenmode with a real eigenfrequency, \textit{i.e.}, a BIC. If this condition is satisfied, the right singular vector $\mathbf{a}^{\rm eff}(\omega,\mathbf{k}_{\parallel})$ corresponding to the eigenfrequency $\omega=\omega_{\rm BIC}$ contains the multipole coefficients $a_{\ell,m}^{\rm eff}(\omega,\mathbf{k}_{\parallel})$. Inserting them into Eq.~\eqref{eq_P} or~\eqref{eq_P_zonal} must give us that the far-field radiation amplitudes for the BIC are $\mathcal{P}_{+} = \mathcal{P}_- = 0$. It is worth mentioning that criteria~\eqref{eq_det} and~\eqref{eq_smin} are directly applicable to optical BICs with the corresponding change of the effective T-matrix.

\subsection{Even BIC in the monopole-quadrupole approximation}
Equation~\eqref{eq_smin} can provide us with resonance frequencies of BICs in arbitrary systems. Here, we give an example of the matrix $\mathbf{T}^{-1}_{\rm eff}(\omega,\mathbf{k}_{\parallel})$ for which Eq.~\eqref{eq_smin} has a solution with $\mathbf{k}_{\parallel} = \bm{0}$ and a real eigenfrequency. For this mode, acoustic radiation in the $\pm z$ directions is suppressed, so that the metasurface does not radiate at all, and the mode can be referred to as an accidental BIC at the $\Gamma$-point. According to selection rules~\eqref{eq_selection_rules}, eigenmodes (including BICs) of a lattice of spheres can have either even or odd parity, which also determines their multipolar content. In the following, we choose the case of even parity. However, the odd-parity modes, which can also enable an accidental BIC, can be considered in the same manner (see Appendix~\ref{sec:odd}).

\begin{figure}
    \centering
    \includegraphics[scale=0.58]{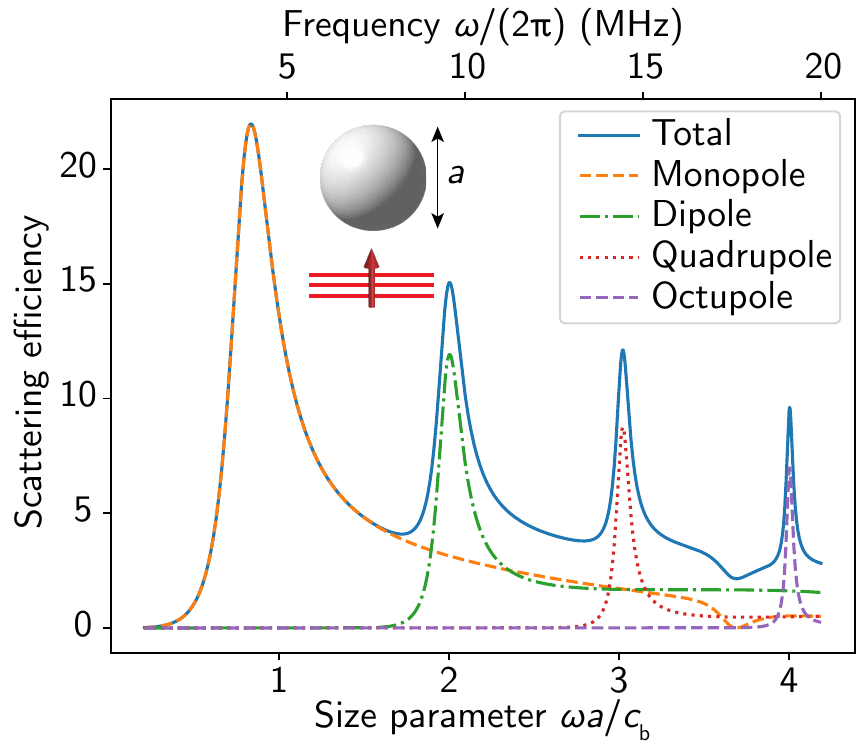}
    \caption{Scattering efficiency of a spherical particle as a function of its size parameter and frequency (blue solid) and contributions to scattering from the monopole (dashed orange), dipole (dashed-dotted green), quadrupole (dotted red), and octupole (dashed violet). The maximum multipole degree is $\ell_{\rm max} = 6$, and the other parameters are listed in Sec.~\ref{sec:simulation_lattice}.}
    \label{fig:xs}
\end{figure}

Let us consider an eigenmode with even parity of the square lattice [$\mathfrak{n}_{\rm L} = 4$ in Eq.~\eqref{eq_selection_rules}] of spheres, for which the following multipole coefficients can be nonzero:
\begin{align}
\label{eq_even_multipoles}
    a_{0,0}^{\rm eff}; a_{2,0}^{\rm eff}; a_{4,0}^{\rm eff}; a_{4,\pm4}^{\rm eff}; ...; a_{2\ell',4m'}^{\rm eff}; ...\, ,
\end{align}
where $\ell' \in \mathds{N}_0$ and $m' \in \mathds{Z}$, such that $|2m'| \leq \ell'$. The other multipole contributions must be zero according to selection rules~\eqref{eq_selection_rules}. Furthermore, we also consider spherical scatterers (resonators) whose acoustic response in the given frequency range can be sufficiently described by the zonal monopole (M), dipole (D), quadrupole (Q), and octupole (O). At the same time, higher-degree multipoles can be neglected (see Fig.~\ref{fig:xs}).
In this case, we can keep only two terms from Eq.~\eqref{eq_even_multipoles}, the zonal monopole $a_{\rm M}^{\rm eff} \equiv a_{0,0}^{\rm eff}$ and the zonal quadrupole $a_{\rm Q}^{\rm eff} \equiv a_{2,0}^{\rm eff}$. For $\mathbf{k}_{\parallel} = \bm{0}$, all unit cells in the array will have identical zonal monopole $a_{\rm M}^{\rm eff}$ and zonal quadrupole $a_{\rm Q}^{\rm eff}$ moments, according to Eq.~\eqref{eq_bloch}. In Eq.~\eqref{eq_mse_lattice_2}, the parts of the single-particle T-matrix and the lattice sum matrix corresponding to the zonal monopole and the zonal quadrupole read as
\begin{align}
    \mathbf{T}(\omega) = 
    \begin{pmatrix}
        T_{\rm MM}& T_{\rm MQ} \\
        T_{\rm QM}& T_{\rm QQ}
    \end{pmatrix}\, , \quad\quad  \mathbf{\Sigma}'(\omega) = 
    \begin{pmatrix}
        \Sigma'_{\rm MM}& \Sigma'_{\rm MQ} \\
        \Sigma'_{\rm MQ}& \Sigma'_{\rm QQ}
    \end{pmatrix}\, ,
\end{align}
where $T_{\rm MM},~T_{\rm MQ},~T_{\rm QM},$ and $T_{\rm QQ}$ [resp. $\Sigma'_{\rm MM} \equiv \Sigma'_{\rm 0,0,0,0}(\omega,\bm{0}),~ \Sigma'_{\rm MQ} \equiv \Sigma'_{\rm 0,0,2,0}(\omega,\bm{0}),$ and $\Sigma'_{\rm QQ} \equiv \Sigma'_{\rm 2,0,2,0}(\omega,\bm{0})$] are couplings between the corresponding multipoles induced by the particle [resp. lattice]. The definition of $\Sigma'_{\ell,m,\ell',m'}(\omega,\mathbf{k}_{\parallel})$ is given by Eq.~\eqref{eq_lattice_sums}. An isolated sphere does not possess particle-induced coupling; therefore, $T_{\rm MQ} = T_{\rm QM} = 0$. Therefore, the matrix $\mathbf{T}^{-1}_{\rm eff}(\omega)$ from Eq.~\eqref{eq_mse_lattice}, when only the even multipoles are considered, becomes
\begin{align}
\label{eq_A}
    \mathbf{T}^{-1}_{\rm eff}(\omega) = 
    \begin{pmatrix}
        T_{\rm MM}^{-1} -  \Sigma'_{\rm MM} & -\Sigma'_{\rm MQ} \\
        -\Sigma'_{\rm MQ} & T_{\rm QQ}^{-1}  - \Sigma'_{\rm QQ}
    \end{pmatrix}\, ,
\end{align}
where, for a sphere, $T_{\rm MM} \equiv \mathfrak{a}_0$ and $T_{\rm QQ} \equiv \mathfrak{a}_2$ (see Appendix~\ref{sec:tm_sphere} for the definition of $\mathfrak{a}_\ell$).
Thus, for a biperiodic lattice of spheres, the monopole-quadrupole coupling is fully induced by the lattice. It can be considered as a channel that enables the formation of the BIC in the symmetric metasurface with two subsystems, the sublattices of monopoles and quadrupoles.

For even zonal multipoles, $\mathcal{P}_{+} = \mathcal{P}_{-}$ according to Eq.~\eqref{eq_P_zonal}. The formation of a BIC corresponds to $\mathcal{P}_{+} = \mathcal{P}_{-} = 0$, which implies the following relationship between the multipole moments of the \textit{even} BIC:
\begin{align}
\label{eq_criterion1}
     a^{\rm eff}_{\rm M} -  \sqrt{5} a^{\rm eff}_{\rm Q} =0\, .
\end{align}
Condition~\eqref{eq_criterion1} ensures that the eigenmode is nonradiant (trapped), \textit{i.e.}, a BIC. The nontriviality of the solution $a^{\rm eff}_{\rm M} \neq 0$ and $a^{\rm eff}_{\rm Q}\neq 0$ is guaranteed by condition~\eqref{eq_smin} [or~\eqref{eq_det}] being satisfied.  We also recall that identity~\eqref{eq_criterion1} is equivalent to solving Eq.~\eqref{eq_lattice_modes} with~\eqref{eq_A} under the assumption that $L < \lambda_{\rm b}$ (see Appendix~\ref{sec:conditions}).

The BIC formation in this system occurs when two radiating multipoles (in our case, zonal) with the same parity interfere, creating a nonradiant state. This mechanism is accessible for an acoustic resonator on an infinite lattice only when acoustic radiation must be suppressed solely in the normal directions. Since the radiation patterns of the multipoles do not overlap [see Fig.~\ref{fig:sketch}(b)], it is impossible in this case to suppress the omnidirectional acoustic radiation for a single resonator through destructive interference between two multipoles. Moreover, BIC formation in a metasurface via this mechanism is possible only when at least two zonal multipoles are taken into account. In contrast, a single-multipole approximation is insufficient to obtain a nontrivial solution, because a single zonal multipole always radiates acoustic waves along its axis. 

It is worth mentioning that the considered metasurface can also support an \textit{odd} BIC formed due to destructive interference between the dipole and the octupole that satisfy the following constraint (see Appendix~\ref{sec:odd}) 
\begin{align}
\label{eq_criterion_odd}
    a^{\rm eff}_{\rm D} -\sqrt{7/3} a^{\rm eff}_{\rm O}  = 0.
\end{align}
The odd BIC emerges close to the frequency of the octupole resonance of a single scatterer, similar to the even BIC, which occurs close to the quadrupole resonance of a single scatterer.

Finally, let us discuss the impact of lattice and resonator symmetries. The form of matrix~\eqref{eq_A} remains the same for the hexagonal lattice [$\mathfrak{n}_{\rm L} = 6$ in Eq.~\eqref{eq_selection_rules}], but with different values of the lattice sums. The monoclinic and orthorhombic lattices ($\mathfrak{n}_{\rm L} = 2$) enable the coupling between the zonal multipoles and the multipoles with $m = \pm 2$, which, however, do not radiate acoustic waves in the normal direction [see Eq.~\eqref{eq_P_zonal}]. For a nonspherical resonator on a metasurface, the eigenmodes, including BICs, can be classified by parity as long as the metasurface possesses inversion symmetry. In its absence, the zonal even and odd multipoles are coupled, and the genuine BICs with even or odd parity transform into quasi-BICs for the parameters when Eq.~\eqref{eq_criterion1} or~\eqref{eq_criterion_odd} is satisfied, respectively (small shape asymmetries of the resonators, which can occur in reality, will also lead to a decrease in the radiative $Q$ factor). To establish a genuine BIC in a metasurface without inversion symmetry, the multipole moments must simultaneously satisfy Eqs.~\eqref{eq_criterion1} and~\eqref{eq_criterion_odd}. In contrast, breaking rotational symmetry of the resonator with respect to the $z$ axis while preserving inversion symmetry induces only coupling between zonal and nonzonal multipoles with the same parity~\cite{Tsimokha2022Apr}, which does not affect Eqs.~\eqref{eq_criterion1} and~\eqref{eq_criterion_odd}; therefore, genuine BICs with a certain parity can exist in such metasurfaces. Note that additional off-diagonal terms due to $T_{\rm MQ} = T_{\rm QM} \neq 0$ appear in Eq.~\eqref{eq_A} for nonspherical particles.

\subsection{Realization in an infinite metasurface}\label{sec:simulation_lattice}

We consider spheres of diameter $a = 50$ $\mu$m made of material with density $\rho_{\rm s} = 360$ kg/m$^3$ and speed of sound $c_{\rm s} = 600$ m/s corresponding to aerogel in the ultrasound range~\cite{Xie1998Feb}. The background medium has the parameters $\rho_{\rm b} = 998$ kg/m$^3$ and $c_{\rm b} = 1500$ m/s that correspond to water at room temperature. For now, material losses are absent. Note that the considered effects can potentially be demonstrated with other high-index acoustic scatterers, for example, air bubbles in water~\cite{Kafesaki2000Jun,Leroy2016Dec,Horvath2023Oct,Kyrimi2024Oct}, or labyrinthine resonators in air~\cite{Zhang2022Apr}. The spherical scatterers are arranged in a square lattice with a lattice constant of $L$, which is the same in both the $x$ and $y$ directions [see Fig.~\ref{fig:sketch}(a)]. The T-matrix of the isolated sphere, as well as the metasurface response, is simulated using our open-source code \textit{acoustotreams} (v.0.2.5) available as a Python package~\cite{acoustotreams}. The lattice sums $\boldsymbol{\Sigma}'$ are efficiently and accurately calculated using Ewald`s summation technique~\cite{Beutel2023Jan} numerically implemented in the subpackage \textit{treams.lattice}~\cite{Beutel2024Apr}. For a sphere diameter of 50 $\mu$m, we consider the ultrasound frequency range, namely, 1-20 MHz. This leads to a size parameter of $\omega a/c_\mathrm{b}$ ranging from 0.21 to 4.19. In Fig.~\ref{fig:xs}, one can see that the scatterer exhibits a resonant response that can be approximated in this spectral range by the first four multipoles, \textit{i.e.}, $\ell_{\rm max} = 3$. We note that we can alternate the frequency range by varying the scatterer size $a$ or the background material $c_{\rm b}$ due to the scaling character of the problem as $\omega a / c_{\rm b}$. 

 \begin{figure*}
    \centering
    \includegraphics[scale=0.49]{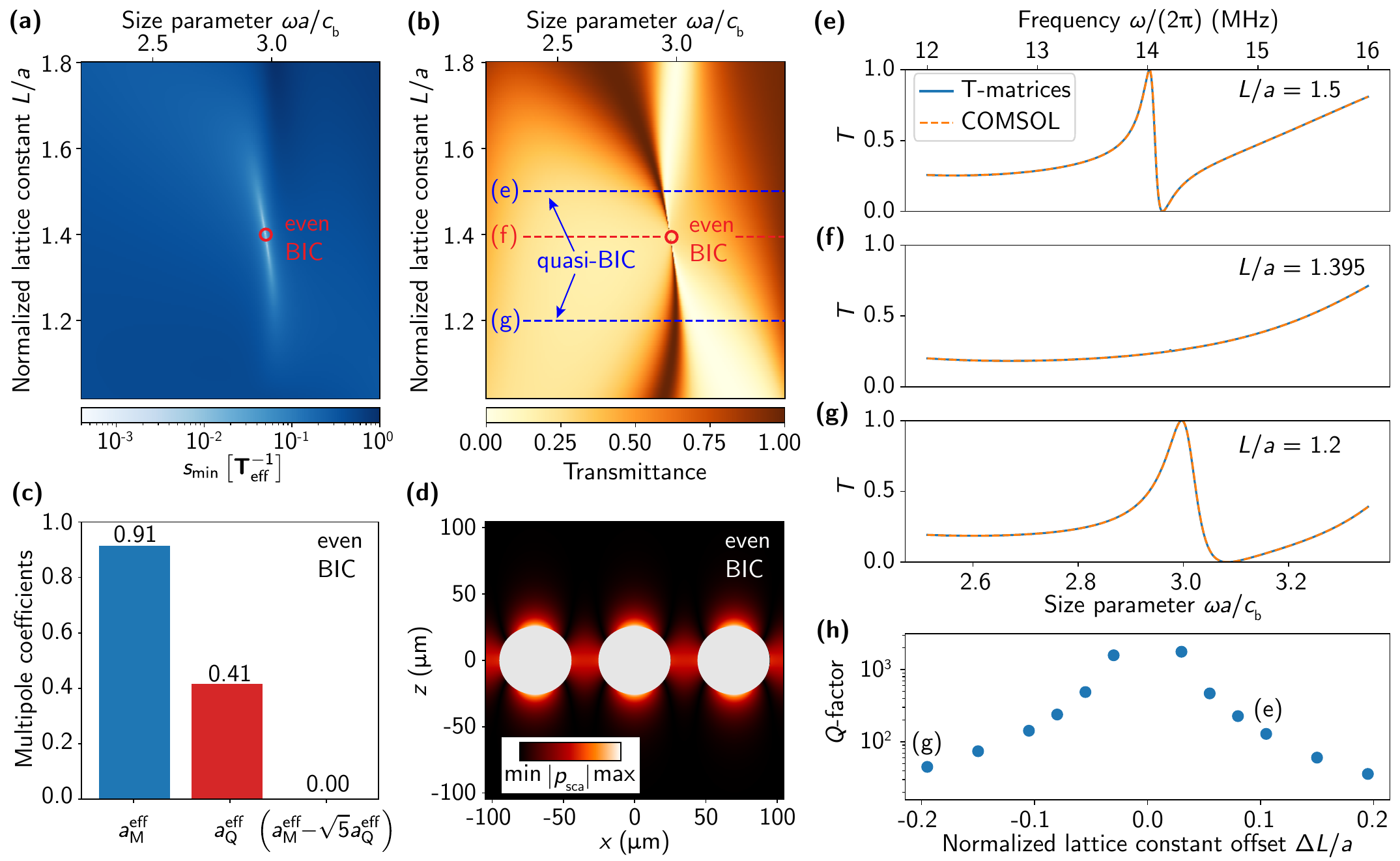}
    \caption{Even BIC in the metasurface of spherical resonators depicted in Fig.~\ref{fig:sketch}(a). (a) The lowest singular value of the inverse effective T-matrix in the monopole-quadrupole approximation~\eqref{eq_A}.  (b) Transmittance of the metasurface $T \equiv |t|^2$ for a normally incident pressure plane wave with $\ell_{\rm max} = 6$ in Eq.~\eqref{eq_t}. The quantities in panels (a) and (b) are plotted as a function of the normalized lattice constant $L/a$ and size parameter $\omega a / c_{\rm b}$, where $a = 50$ $\mu$m and $c_{\rm b} = 1500$ m/s (the other parameters are listed in the text). The red circles indicate the lattice constant and the frequency for which the even BIC appears. (c) Effective zonal monopole and quadrupole coefficients for the even BIC. (d) Normalized absolute value of the pressure field generated by the BIC in the $xz$ plane outside the spherical resonators. (e-g) Transmittance of the metasurface as a function of frequency for normalized lattice constants [panel (e)] $L/a = 1.5$, [panel (f)] $L/a = 1.395$, and [panel (g)] $L/a = 1.2$. The solid blue and dashed orange curves correspond to the values computed by the T-matrix method (in \textit{acoustotreams}) and the finite element method (\textit{pressure acoustics,frequency domain} interface in COMSOL Multiphysics\texttrademark, \ v. 5.5), respectively. (h) $Q$ factor of the quasi-BIC resonance as a function of the normalized lattice constant offset $\Delta L/a = (L/a-L_{\rm BIC}/a)$ where $L_{\rm BIC}/a = 1.395$.}
    \label{fig:lattice_spheres}
\end{figure*}

To perform the singular decomposition of matrix~\eqref{eq_A}, we employ the function \texttt{numpy.linalg.svd} in Python 3.11. Figure~\ref{fig:lattice_spheres}(a) shows the lowest singular value of matrix~\eqref{eq_A} as a function of the normalized lattice constant $L/a$ and the size parameter $\omega a / c_{\rm b}$. One can see that the value tends to zero for $\omega a / c_{\rm b} = 2.975$ and $L/a = 1.395$ [respectively, $\omega/(2\pi) = 14.2$ MHz and $L = 69.75$ $\mu$m], indicated by the red marker. Thus, the lattice response is resonant for a real-valued frequency, as Eq.~\eqref{eq_smin} dictates. Hence, this eigenmode is supposed to correspond to the even BIC. Indeed, Fig.~\ref{fig:lattice_spheres}(c) depicts the effective monopole and quadrupole moments of a unit cell for the BIC obtained as the right singular vector corresponding to the zero singular value of Eq.~\eqref{eq_A}. One can clearly see that the moments obey condition~\eqref{eq_criterion1}, which guarantees the absence of acoustic radiation. Moreover, each moment separately is not equal to zero since Eqs.~\eqref{eq_det} and~\eqref{eq_smin} are satisfied. Therefore, the pressure field of the BIC is not zero, but is trapped near the structure, as shown in Fig.~\ref{fig:lattice_spheres}(d). Indeed, although the intensity of the radiated pressure field in the far-field region is zero ($\mathcal{P}_{\pm} = 0$), the near field corresponding to the BIC is nonzero, since evanescent multipole fields of different degrees decay differently when moving away from the source.

Furthermore, the resonance associated with the BIC can also be observed in the transmittance spectrum for a normally incident pressure plane wave $p_{\rm inc}(\mathbf{r}) = \mathrm{e}^{\mathrm{i} kz}$. To find the metasurface response, we calculate effective multipole moments~\eqref{eq_a_eff} up to $\ell_{\rm max} = 6$ and then calculate transmission~\eqref{eq_t} using \textit{acoustotreams}. Figure~\ref{fig:lattice_spheres}(b)
displays the transmittance $T\equiv|t|^2$ of the metasurface as a function of the normalized lattice constant and size parameter, while Figs.~\ref{fig:lattice_spheres}(e)-(g) plot it as a function of the size parameter for the different values of the normalized lattice constant. In Fig.~\ref{fig:lattice_spheres}(b), one can see that the transmittance exhibits a resonant feature associated with the BIC for the same lattice constant and frequency where the lowest singular value vanishes. Note that exactly at the frequency of the BIC formed by the resonant monopole and quadrupole, the resonance disappears from the spectrum because the incident wave cannot excite a nonradiant mode, and the transmittance is therefore entirely determined by the nonresonant multipoles (dipole and octupole) [see also Fig.~\ref{fig:lattice_spheres}(f)]. However, as shown in Figs.~\ref{fig:lattice_spheres}(e) and~\ref{fig:lattice_spheres}(g), one can observe the development of Fano-type resonances (quasi-BICs) with relatively high quality factors ($Q$ factors) by gradually varying the lattice constant. Note that the Fano profile of the considered resonances arises from the interference (overlapping) of the incident field and the fields of the even multipoles (monopole and quadrupole) from the resonant quasi-BIC mode, with the addition of odd multipoles (dipole and octupole) from the nonresonant mode~\cite{allayarov2024anapole}.  To estimate the $Q$-factor of a Fano-type profile of the quasi-BIC, we fit the transmittance spectrum for a given lattice constant with a sum of several Fano resonances as described in Ref.~\onlinecite{Mikhailovskii2024Apr}. In Figure~\ref{fig:lattice_spheres}(h), the $Q$-factor decreases as the inverse square of the normalized lattice constant offset $|\Delta L|/a \equiv |L - L_{\rm BIC}|/a$, where the normalized lattice constant of the BIC is $L_{\rm BIC}/a = 1.395$, since Eq.~\eqref{eq_criterion1} ceases to hold when $|\Delta L|/a > 0$. Note that the $Q$-factor of the genuine BIC at $L_{\rm BIC}/a$ is infinite. Finally, to verify the transmittance calculated with the T-matrix method, we also compare it in Figs.~\ref{fig:lattice_spheres}(e)-(g) with that computed by the pressure acoustics module of COMSOL Multiphysics\texttrademark,  v.5.5~\cite{comsol}. Appendix~\ref{sec:fem} concisely describes the model. The figures confirm that both methods demonstrate perfect agreement.

\section{Quasi-BIC regime}\label{sec:quasi-bic}

As shown in the previous section, infinite lattices of spherical particles can support acoustic BICs resulting from multipole interference at specific frequencies and lattice constants. By tuning the lattice constant, the BIC with an infinite $Q$ factor can be transformed into a quasi-BIC, which manifests itself in the spectrum as a Fano resonance with a large $Q$ factor compared to resonances of an isolated resonator. In practice, the quasi-BIC can be more important in applications than the genuine BIC, since it can be excited by a normally incident plane wave. The acoustic quasi-BICs have already been utilized to enhance the acoustic Purcell factor~\cite{Landi2018Mar,Huang2024Jan}, and to achieve perfect absorption~\cite{Krasikova2024Aug,Zhang2023Aug} or localization of sound energy~\cite{Huang2020Aug}. In this section, we discuss the acoustic quasi-BIC regime under realistic conditions, where either material losses or a substrate is present, and also its manifestation in finite-size arrays.    

The total $Q$ factor of an eigenmode can be decomposed into two terms as 
\begin{align}
\label{eq_Q}
    Q_{\rm tot}^{-1} = Q_{\rm rad}^{-1} + Q_{\rm nr}^{-1}\, ,
\end{align}
where $Q_{\rm rad}$ and $Q_{\rm nr}$ are responsible for radiative and nonradiative (material) losses, respectively. For a genuine BIC in the infinite lattice, $Q_{\rm rad}^{-1} = 0$ and consequently the $Q$ factor is limited by $Q_{\rm tot} = Q_{\rm nr}$ even for a genuine BIC~\cite{Koshelev2023May}. For a quasi-BIC in both finite and infinite structures, $Q_{\rm rad}^{-1} > 0$ and Eq.~\eqref{eq_Q} should be used to calculate the $Q$ factor.

\subsection{Influence of material absorption and substrate}

\begin{figure}
    \centering
    \includegraphics[scale=0.56]{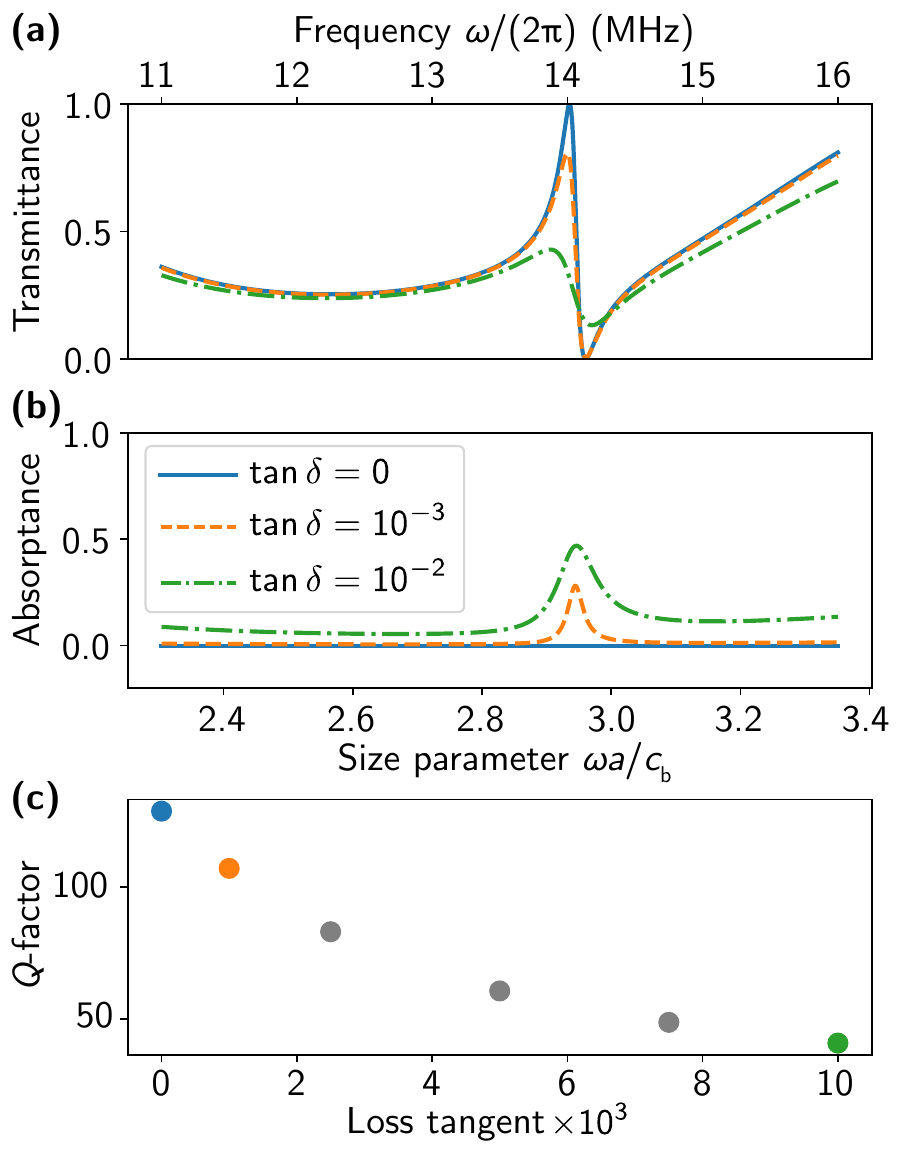}
    \caption{Influence of material losses in scatterers (resonators) on the quasi-BIC resonance. (a) Transmittance of the metasurface $|t|^2$ [see Eq.~\eqref{eq_t}] in the quasi-BIC regime for different values of material loss tangent being 0 (blue), $10^{-3}$ (orange), and $10^{-2}$ (green). The normalized lattice constant is $L/a = 1.5$. (b) The corresponding absorptance $(1-|r|^2 - |t|^2)$ of the metasurface. (c) Total $Q$ factor of the quasi-BIC as a function of material loss tangent.}
    \label{fig:T_sphere_abosrption}
\end{figure}

First, we assume that the particles have material losses by considering the complex-valued acoustic impedance $Z_{\rm s} = Z_{\rm s}^{\prime}+\mathrm{i}Z_{\rm s}^{\prime\prime}$, where $Z_{\rm s}'= \rho_{\rm s} c_{\rm s}= 0.216$ MRayl.  The ratio of the imaginary part to the real part defines the so-called loss tangent $\tan \delta = Z_{\rm s}^{\prime\prime}/Z_{\rm s}^{\prime}$. Figure~\ref{fig:T_sphere_abosrption}(a) shows the transmittance spectra of the metasurface with the normalized lattice constant of $L/a = 1.5$ for different values of the loss tangent, while Fig.~\ref{fig:T_sphere_abosrption}(b) shows the corresponding absorptance spectra. Figure~\ref{fig:T_sphere_abosrption}(c) plots the values of the total $Q$ factor of the quasi-BIC, which are estimated using the procedure from Ref.~\onlinecite{Mikhailovskii2024Apr} to fit the resonances in Fig.~\ref{fig:T_sphere_abosrption}(a). For the resonators without material losses, the quasi-BIC has the radiative $Q$ factor and consequently the total  $Q$ factor of $Q_{\rm tot} = Q_{\rm rad} \approx 129$, whereas the absorptance is zero [see the blue curves in Figs.~\ref{fig:T_sphere_abosrption}(a) and~\ref{fig:T_sphere_abosrption}(b)]. Meanwhile, one can observe that, as the material losses increase, the total $Q$ factor decreases [Fig.~\ref{fig:T_sphere_abosrption}(c)], the absorptance at the quasi-BIC frequency increases [Fig.~\ref{fig:T_sphere_abosrption}(b)], and the resonance in the transmittance spectrum weakens [Fig.~\ref{fig:T_sphere_abosrption}(a)]. Thus, an acoustic quasi-BIC resonance in a lossy system appears as an absorptance resonance, as well as its electromagnetic counterpart~\cite{allayarov2024anapole}. Moreover, for the lossy resonators, the decrease in the nonradiative $Q$ factor with increasing loss tangent can be justified by its approximate upper-bound expression, $Q_{\rm nr} = 1/\tan \delta$~\cite{Mikhailovskii2024Apr}.

\begin{figure}[h!]
    \centering
    \includegraphics[scale=0.57]{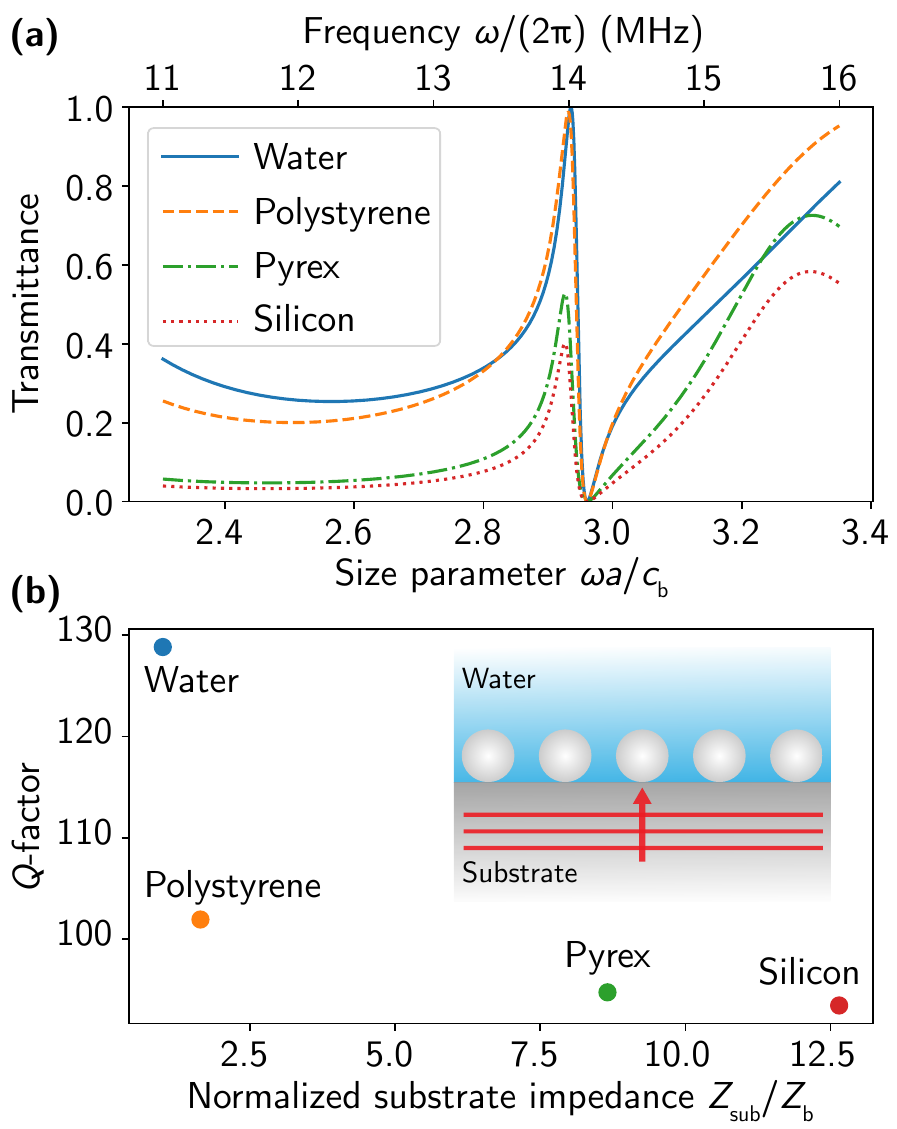}
    \caption{Influence of a substrate on the quasi-BIC resonance. (a) Transmittance of the metasurface $|t|^2$ [see Eq.~\eqref{eq_t}] in the quasi-BIC regime for different substrates. The normalized lattice constant is $L/a = 1.5$. (b) $Q$ factor of the quasi-BIC as a function of the normalized substrate impedance $Z_{\rm sub}/Z_{\rm b}$ with $Z_{\rm b} = 1.497$ MRayl.}
    \label{fig:T_sphere_substrate}
\end{figure}

Similarly, we can analyze the impact of a substrate. Since the considered BIC emerges for the specific values of the parameters, it is sensitive to symmetry breaking, as well as other accidental BICs~\cite{Koshelev2023May}. Hence, the larger the impedance contrast between the upper and lower half-spaces, the smaller the $Q$ factor. Figure~\ref{fig:T_sphere_substrate}(a) shows the transmittance spectra of the metasurface with the normalized lattice constant of $L/a = 1.5$ for different substrates. The material parameters of the substrates are taken from Ref.~\onlinecite{Bruus2011Dec}. Figure~\ref{fig:T_sphere_substrate}(b) plots the $Q$ factor of the resonance as a function of the normalized substrate impedance $Z_{\rm sub}/Z_{\rm b}$, where $Z_{\rm b} = \rho_{\rm b} c_{\rm b} = 1.497$ MRayl is the impedance of the upper half-space (note that $Q_{\rm nr}^{-1} = 0$). One can clearly see that if the impedance contrast $Z_{\rm sub}/Z_{\rm b}$ increases, the $Q$ factor decreases. Thus, acoustic metasurfaces in the quasi-BIC regime are strongly responsive to background variations and can therefore serve as the basis for highly sensitive acoustic sensors.

\subsection{Realization in finite-size arrays}

The considered BIC has an infinite $Q$ factor because translation symmetry of the lattice allows for only one radiation channel, being the $z$ axis, whereas the scattering amplitudes~\eqref{eq_P_zonal} for the directions $-z$ and $+z$ are zero for the BIC due to Eq.~\eqref{eq_criterion1}. Therefore, the overall radiation of the lattice eigenmode is zero, and the mode is trapped (BIC). In the experiment, all structures have finite sizes along all dimensions. Since a finite-size array has an infinite number of allowed radiation channels, it always possesses nonzero radiation losses~\cite{Silveirinha2014Feb,Hsu2016Jul, Koshelev2023May}. Thus, finite-size structures can possess neither electromagnetic nor acoustic BICs with an infinite $Q$ factor. Note that elastic spherical resonators, which support shear waves, can possess a genuine transversely polarized BIC due to the polarization orthogonality between the fields inside and outside the resonator~\cite{Deriy2022Feb}, but we do not consider them in this work. Thus, if a lattice of acoustic resonators becomes a finite-size $N \times N$ array, the considered BIC always transforms into a quasi-BIC. Moreover, the response of finite-size arrays can remarkably differ from that of an infinite lattice~\cite{Zakomirnyi2019Dec,Rodriguez2012Oct,Ustimenko2024Mar}. Hence, we also discuss the development of the BIC resonance for finite arrays. 

\begin{figure}
    \centering
    \includegraphics[scale=0.53]{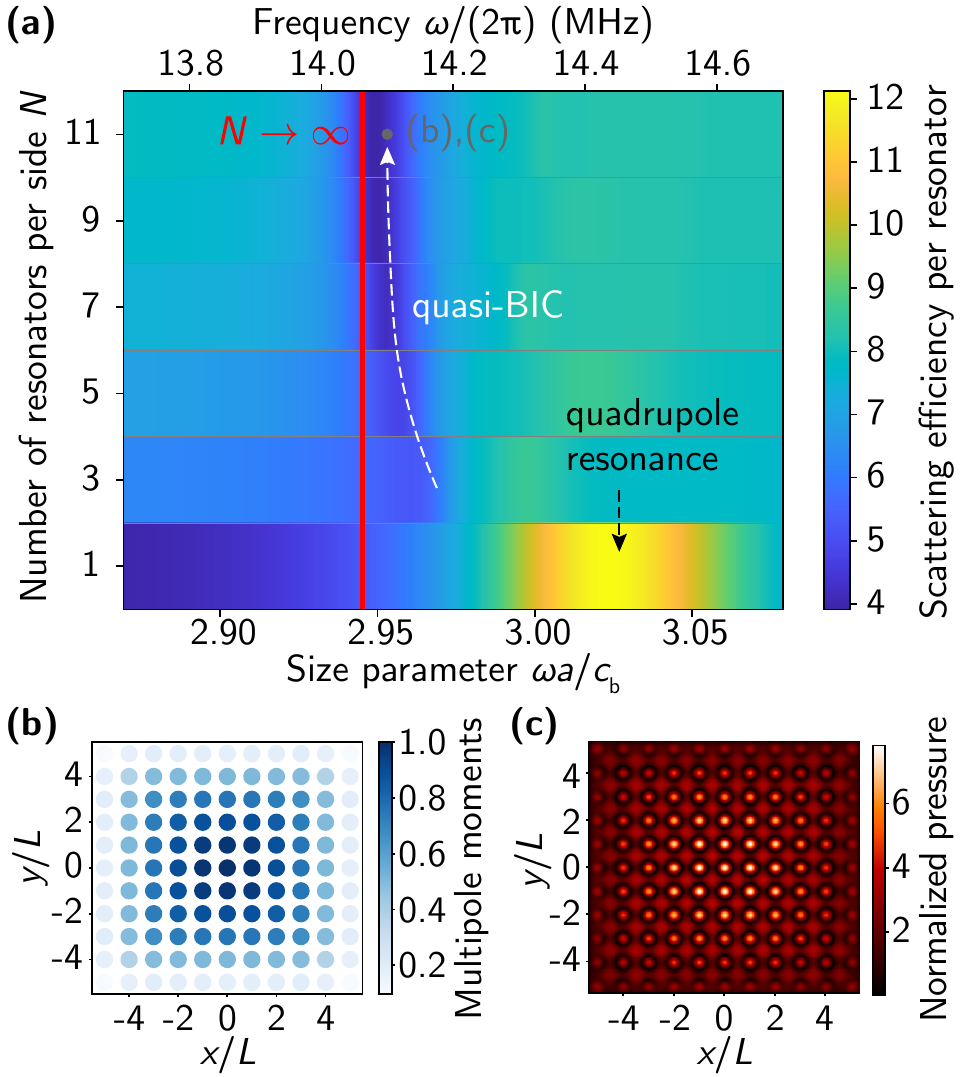}
    \caption{Effective scattering efficiency per resonator for an $N \times N$ array of spherical resonators with the normalized lattice constant of $L/a = 1.5$ at normal incidence ($\ell_{\rm max} = 3$). The red line indicates the frequency of the quasi-BIC resonance of the infinite lattice with the same lattice constant. (b)-(c) The quasi-BIC resonance in the $11 \times 11$ array at $\omega a/c_{\rm b} = 2.95$ ($\lambda_{\rm b}/a = 1.59$). (b) Normalized multipole content of the resonators $\sum_{\ell,m}|a_{i,\ell,m}|^2/\max_i \sum_{\ell,m}|a_{i,\ell,m}|^2$. (c) The scattered pressure field distribution in the near field region in $z = 30$ $\mu$m ($ \approx 0.4\lambda_{\rm b}$)  plane.}
    \label{fig:finite-size}
\end{figure}

To simulate the acoustic response of finite-size arrays, we also use the T-matrix-based formalism. Within the approach, the scattered pressure field of the $i$th sphere is again described by a vector of multipole coefficients $\mathbf{a}(\mathbf{r}_i)$ as well as in the infinite-lattice case. However, because of the broken translation symmetry, the coefficients are no longer linked through the Bloch theorem, but via a system of linear equations considering mutual coupling between resonators (see Appendix~\ref{sec:Tmatrix_finite}). A solution to system~\eqref{eq_mse}, which can be obtained using \textit{acoustotreams} code, provides us with the self-consistent response of the finite-size array. We can also calculate the scattering cross section of the $N \times N$ array $\sigma_N(\omega)$ and divide it by $N^2\sigma_{\rm  geom}$ to obtain the effective scattering efficiency of a particle in the array. In Fig.~\ref{fig:finite-size}(a), it is plotted as a function of frequency and $N$ when the normalized lattice constant is fixed at the value of $L/a=1.5$ corresponding to the quasi-BIC resonance in the infinite metasurface [see Fig.~\ref{fig:lattice_spheres}(e)]. First, it is notable that even the response of the $3 \times 3$ array differs significantly from that of the single scatterer. Namely, the quasi-BIC resonance appears already for $N = 3$, corresponding to a minimum in the scattering efficiency. Moreover, as $N$ increases, the resonance frequency approaches the infinite-array limit and converges by $N = 11$.

Figures~\ref{fig:finite-size}(b) and~\ref{fig:finite-size}(c) show the multipole moments of the resonators on the $11 \times 11$ array and the pressure field, generated by this array in the near-field region. One can see that the farther a resonator is from the center, the lower the near field is above it and the weaker the multipole moments are induced in the resonator, forming approximately a standing-wave pattern between the edges of the array with the wave vector $k_x = k_y = \frac{\pi}{(N+1)L}$ [see Eq.~(2) in Ref.~\onlinecite{Volkov2024Feb}]. Thus, a normally incident plane wave mostly excites resonators that are close to the center, yielding strong (enhanced) near fields. In particular, it implies that the quasi-BIC should be better coupled to an external source that is located above the array center rather than above its edges.

\section{Conclusion}
In this work, we investigate acoustic bound states in the continuum (trapped modes) in biperiodic metasurfaces with one acoustic resonator (scatterer) per unit cell. First, we develop a comprehensive theoretical model based on the T-matrix method and symmetry analysis that enables us to obtain the analytical conditions of the BIC formation. Next, we show that an accidental acoustic BIC can emerge at the $\Gamma$-point of a metasurface with inversion symmetry due to destructive interference between two effective zonal multipoles of the same parity. Our numerical simulations reveal a spectral feature associated with the BIC in the spectrum of the metasurface when a pressure plane wave is normally incident. Furthermore, a slight detuning of the lattice constant transforms the genuine BIC into a quasi-BIC, which manifests in the spectrum as a Fano resonance.

Focusing on the experiment, we also consider the quasi-BIC regime under realistic assumptions, including either material losses in the resonators or a substrate. Moreover, we investigate the development of the quasi-BIC resonance in finite $N \times N$ arrays of particles as a function of $N$ and identify the values of $N$ sufficient for convergence to the infinite-lattice limit. 

Our results not only provide the theoretical and numerical foundations for acoustic BICs in metasurfaces with inversion symmetry but also demonstrate their practical feasibility under realistic conditions. The ability to control the transition from genuine BICs to quasi-BICs through simple geometric tuning provides a versatile mechanism for engineering highly selective, low-loss acoustic responses. This approach opens opportunities for designing advanced acoustic metasurfaces with tunable functionalities. Potential applications range from ultranarrowband acoustic filters and high-sensitivity sensors to energy-harvesting devices and medical ultrasound technologies. By bridging fundamental physics and practical design strategies, our work paves the way for next-generation acoustic systems that exploit the unique properties of BICs for enhanced wave control and device performance.
 
 \begin{acknowledgments}
The authors thank Kristina Frizyuk, Puneet Garg, Oscar Demeulenaere, Athanasios Athanassiadis, and Peer Fischer for fruitful discussions. N.U. and C.R. acknowledge support through the Deutsche Forschungsgemeinschaft (DFG, German Research Foundation) under Germany’s Excellence Strategy via the Excellence Cluster 3D Matter Made to Order (EXC-2082/1, Grant No. 390761711) and from the Carl Zeiss Foundation via CZF-Focus@HEiKA. A.B.E. acknowledges support of the Deutsche Forschungsgemeinschaft (DFG, German Research Foundation) under Germany’s Excellence Strategy within the Cluster of Excellence PhoenixD (EXC 2122, Project ID No. 390833453).
\end{acknowledgments}

\section*{DATA AVAILABILITY}
The data that support the findings of this paper are openly
available~\cite{acoustotreams}.

 \appendix
 \section{Scalar spherical waves and their translation coefficients}\label{sec:ssw}
The scalar spherical wave (multipole) of degree $\ell=0,1,2,...$ and order $m= -\ell, -\ell + 1, ..., \ell - 1, \ell$ is defined as
\begin{align}
\label{eq_psi}
    \Psi^{(n)}_{\ell, m}(\omega; \mathbf{r}) = z^{(n)}_{\ell}(\omega r/c)Y_{\ell,m}(\theta,\varphi)\, ,
\end{align}
where the auxiliary index is used to distinguish the regular waves of $n = 1$, which are finite at $\mathbf{r} = \bm{0}$, from the singular ones of $n = 3$, which are singular for $\mathbf{r} \to \bm{0}$ and obey the Sommerfeld radiation condition for $r \to +\infty$. The radial dependence is described by the spherical Bessel function of the first kind $z^{(1)}_{\ell}(\omega r/c) = j_{\ell}(\omega r/c)$, or the spherical Hankel function of the first kind $z^{(3)}_{\ell}(\omega r/c) = h^{(1)}_{\ell}(\omega r/c)$. The spherical harmonics are defined by~\cite{Khersonskii1988} 
\begin{align}
\label{eq_Y}
    Y_{\ell, m}(\theta,\varphi) = \sqrt{\frac{2\ell + 1}{4 \pi} \frac{(\ell - m)!}{(\ell + m)!}} P^m_{\ell}(\cos \theta) \mathrm{e}^{\mathrm{i} m \varphi}\, ,
\end{align}
where the associated Legendre polynomials are given by
\begin{align}
\label{eq_Plm}
    P^m_{\ell}(x) = \frac{(-1)^{m}}{2^{\ell}\ell!} (1-x^2)^{m/2} \frac{\mathrm{d}^{\ell + m}}{\mathrm{d}x^{\ell + m}}(x^2-1)^{\ell}.
\end{align}

The scalar spherical multipoles obey the following addition theorem~\cite{Stein1961}:
\begin{align}
    \Psi^{(3)}_{\ell, m}(\omega;\mathbf{r}) = \sum_{\ell',m'} \mathcal{C}^{(3)}_{\ell', m',\ell, m}(\omega; \mathbf{r}-\mathbf{r}_0) \Psi^{(1)}_{\ell', m'}(\omega; \mathbf{r}_0)\, .
\end{align}
The translation coefficients of Eq.~\eqref{eq_psi} are given by
\begin{align}
\label{eq_translation_coeff}
\begin{aligned}
    \mathcal{C}_{\ell', m', \ell, m}^{(n)}(\omega; \mathbf{r}) &= \sqrt{4\pi (2\ell+1)(2\ell'+1)} \ (-1)^{m} \ \mathrm{i}^{\ell'-\ell} \\
    &\times \sum_q \mathrm{i}^{q} \sqrt{2q+1} \Psi^{(n)}_{q,m-m'}(\omega; \mathbf{r}) \\
    &\times \underbrace{\begin{pmatrix}
            \ell& \ell' & q \\
            m & -m' & m'-m
        \end{pmatrix}
        \begin{pmatrix}
            \ell& \ell' & q \\
            0& 0& 0
        \end{pmatrix}}_{\text{Wigner's 3-j symbols}}\, ,
\end{aligned}
\end{align}
where $q \in \left\{\ell'+\ell, \ell'+\ell-2, ..., \max\left(|\ell'-\ell|, |m' - m| \right) \right\}$.

\section{Expansion of a plane wave into scalar spherical waves}\label{sec:plane_wave_exp}
A pressure plane wave $p_{\mathrm{inc}}(\omega;\mathbf{r}) = p_0 \mathrm{e}^{\mathrm{i} \mathbf{k} \cdot \mathbf{r}}$ can be expanded into regular scalar spherical waves~\eqref{eq_psi} as follows:
\begin{align}
    p_{\mathrm{inc}}(\omega;\mathbf{r}) = p_0 \cdot 4\pi \sum\limits_{\ell=0}^{+\infty}  \sum\limits_{m=-\ell}^{m=+\ell} \mathrm{i}^{\ell} Y^*_{\ell, m}(\theta_{\mathbf{k}},\varphi_{\mathbf{k}}) \Psi_{\ell,m}^{(1)}(\omega; \mathbf{r})\, ,
\end{align}
where $\theta_{\mathbf{k}}$ and $\varphi_{\mathbf{k}}$ are the spherical angles of $\mathbf{k}$ with $|\mathbf{k}| = \omega/c$. Hence, the expansion coefficients of the plane wave are $b_{\ell,m} = p_0 4 \pi  \mathrm{i}^{\ell}  Y^*_{\ell, m}(\theta_{\mathbf{k}},\varphi_{\mathbf{k}})$.

\section{Acoustic T-matrix of a sphere}\label{sec:tm_sphere}
For an isolated homogeneous sphere of diameter $a$, the entries of the acoustic T-matrix read as  $T_{\ell,m,\ell',m'} = \mathfrak{a}_{\ell}\delta_{\ell,\ell'} \delta_{m,m'}$, where $\delta_{\ell,\ell'}$ is the Kronecker symbol, and the coefficients $\mathfrak{a}_{\ell}$ are given as~\cite{Toftul2019Oct}
\begin{align}\label{eq_mie_coeffs}
    \mathfrak{a}_{\ell} = \frac{\overline{Z} j'_{\ell}(\eta)j_{\ell}(\xi)-j_{\ell}(\eta)j'_{\ell}(\xi)}{j_{\ell}(\eta) h^{(1)'}_{\ell}(\xi) - \overline{Z} j'_{\ell}(\eta)h^{(1)}_{\ell}(\xi)}\, ,
\end{align}
where $\overline{Z} = \dfrac{c_{\rm b} \rho_{\rm b}}{c_{\rm s} \rho_{\rm s}}$, $\eta = \dfrac{\omega a}{2c_{\rm s}}$, and $\xi = \dfrac{\omega a}{2c_{\rm b}}$. Thus, $T_{\rm MM} = \mathfrak{a}_{0}$ and $T_{\rm QQ} = \mathfrak{a}_{2}$. 

We can rewrite the inverse coefficient in the absence of material losses (absorption), \textit{i.e.}, when the material parameters are real valued, as
\begin{align}
    \frac{1}{\mathfrak{a}_{\ell}} = -1 + \mathrm{i} \underbrace{\frac{j_{\ell}(\eta) y'_{\ell}(\xi) - \overline{Z} j'_{\ell}(\eta)y_{\ell}(\xi)}{\overline{Z} j'_{\ell}(\eta)j_{\ell}(\xi)-j_{\ell}(\eta)j'_{\ell}(\xi)}}_{\in \mathbb{R}}\, .
\end{align}
Here, we have used the definition of the spherical Hankel function $h^{(1)}_{\ell}(\xi) = j_{\ell}(\xi) + \mathrm{i}y_{\ell}(\xi)$, where $y_{\ell}(\xi)$ is the spherical Bessel function of the second kind. Hence, 
\begin{align}
\label{eq_re_inv_mie}
    \Re\left[ \frac{1}{\mathfrak{a}_{\ell}}\right] \equiv -1\,,
\end{align}
if material losses are absent. 

\section{Derivation of Eq.~\eqref{eq_mse_lattice}}\label{sec:lattice}

In a multiple scattering problem, the T-matrix, computed for an isolated object, links the expansion coefficients of the scattered field with those of the so-called local field. The local field for every scatterer is a superposition of the incident and secondary fields generated by other scatterers. For the reference unit cell at $\mathbf{r}_0 = \bm{0}$, the local field is
\begin{align}
\label{eq_ploc_ms}
\begin{aligned}
    &p_{\rm loc}(\omega; \bm{0}) = \sum_{\ell,m} b_{\ell, m} \Psi^{(1)}_{\ell, m}(\omega; \bm{0})  \\ & + \sum_{\ell,m} \sum_{{\bf R} \neq \bm{0}} a^{\rm eff}_{{\bf R},\ell, m} \Psi^{(3)}_{\ell, m}(\omega; - \mathbf{R})\, ,
\end{aligned}
\end{align}
where $a_{{\bf R},\ell, m} (\omega)$ are the expansion coefficients for the scatterer located at $\mathbf{R}$.
For Eq.~\eqref{eq_ploc_ms} to have the same form as Eq.~\eqref{eq_pinc}, we use translation coefficients~\eqref{eq_translation_coeff} to transform the scattered waves into incident ones as
\begin{align}
\label{eq_translation}
    \Psi^{(3)}_{\ell, m}(\omega;-\mathbf{R}) = \sum_{\ell',m'} \mathcal{C}^{(3)}_{\ell', m',\ell, m}(\omega; -\mathbf{R}) \Psi^{(1)}_{\ell', m'}(\omega; \bm{0})\, .
\end{align}
If we substitute Eq.~\eqref{eq_translation} into Eq.~\eqref{eq_ploc_ms}, we can expand the local field at the reference unit cell as
\begin{align}
    p_{\rm loc}(\omega; \bm{0}) = \sum_{\ell,m} \widetilde{b}_{\ell, m}  \Psi^{(1)}_{\ell, m}(\omega; \bm{0})\, ,
\end{align}
with the expansion coefficients
\begin{align}
\label{eq_coeffs_local}
\begin{aligned}
    \widetilde{b}_{\ell,m} = & b_{\ell,m}\\&+\sum_{\ell',m'} \sum_{{\bf R} \neq \bm{0}} \mathcal{C}^{(3)}_{\ell, m,\ell', m'}(\omega; -\mathbf{R}) a_{{\bf R},\ell', m'}^{\rm eff}\, .
\end{aligned}
\end{align}
Substituting Eq.~\eqref{eq_bloch} into Eq.~\eqref{eq_coeffs_local} gives us the following:
\begin{align}
\label{eq_coeffs_local2}
   \widetilde{b}_{\ell, m} = b_{\ell, m} + \sum_{\ell',m'} \Sigma'_{\ell, m,\ell', m'}(\omega, \mathbf{k}_{\parallel}) a_{\ell', m'}^{\rm eff}\, ,
\end{align}
where the lattice sum is defined as
\begin{align}
\label{eq_lattice_sums}
    \Sigma'_{\ell, m,\ell', m'}(\omega, \mathbf{k}_{\parallel}) = \sum_{{\bf R} \neq \bm{0}} \mathcal{C}^{(3)}_{\ell, m,\ell', m'}(\omega; -\mathbf{R})\mathrm{e}^{\mathrm{i} \mathbf{k}_{\parallel}\cdot\mathbf{R}}\, .
\end{align}
Local field coefficients~\eqref{eq_coeffs_local2} can be linked to the scattered field coefficients via the T-matrix as in Eq.~\eqref{eq_Tmatrix}, in which we replace $b_{\ell, m}$ with $\widetilde{b}_{\ell, m}$. Then, we obtain a system of equations to determine $a_{\ell, m}^{\rm eff}$,
\begin{align}
\begin{aligned}
        a_{\ell,m}^{\rm eff} = &\sum_{\ell',m' }T_{\ell,m,\ell',m'}(\omega) \\ &\times \left[ b_{\ell', m'} + \sum_{\ell'',m''} \Sigma'_{\ell', m',\ell'', m''}(\omega, \mathbf{k}_{\parallel}) a_{\ell'', m''}^{\rm eff}\right]\, .
\end{aligned}
\end{align}
If we arrange the coefficients in a vector $\mathbf{a}^{\rm eff}$ in ascending order of $\ell = 0,1,2,...$ and $m=-\ell,-\ell+1,...,\ell$, we obtain Eq.~\eqref{eq_mse_lattice}.

\section{Derivation of the selection rules for $\mathbf{k}_{\parallel} = \bf{0}$}\label{sec:selection_rules}

Let us derive selection rules~\eqref{eq_selection_rules}, \textit{i.e.},  determine nonzero $\Sigma'_{\ell,m,\ell',m'}$ for $\mathbf{k}_{\parallel} = \bf{0}$ given by [see Eq.~\eqref{eq_lattice_sums}]
\begin{align}
\label{eq_lattice_sums_Gamma}
    \Sigma'_{\ell, m,\ell', m'}(\omega, \bm{0}) = \sum_{{\bf R} \neq \bm{0}} \mathcal{C}^{(3)}_{\ell, m,\ell', m'}(\omega; -\mathbf{R})\, .
\end{align}

\subsection*{1. Inversion symmetry}
First, we consider a lattice that has symmetry with respect to the inversion $\mathbf{R} \to -\mathbf{R}$. After the inversion, Eq.~\eqref{eq_lattice_sums_Gamma} becomes the following:
\begin{align}
    \widehat{\Sigma}'_{\ell, m,\ell', m'}(\omega, \bm{0}) = \sum_{{\bf R} \neq \bm{0}} \mathcal{C}^{(3)}_{\ell, m,\ell', m'}(\omega; \mathbf{R})\, .
\end{align}
Since the parity of degree $\ell$ determines the parity of a spherical multipole under the inversion $\Psi^{(n)}_{\ell, m}(\omega; -\mathbf{r}) = (-1)^{\ell} \Psi^{(n)}_{\ell, m}(\omega; \mathbf{r})$, the translation coefficients of scalar spherical waves obey the following property~\cite{Kim2004}:
\begin{align}
    \mathcal{C}^{(3)}_{\ell, m,\ell', m'}(\omega; \mathbf{R}) = (-1)^{\ell+\ell'}\mathcal{C}^{(3)}_{\ell, m,\ell', m'}(\omega; -\mathbf{R})\, .
\end{align}
However, since the lattice remains the same upon inversion, the lattice sums must be equal $\widehat{\Sigma}'_{\ell, m,\ell', m'} = \Sigma'_{\ell, m,\ell', m'}$, which implies that $(\ell+\ell')$ is an even number. This can be true only if $\ell$ and $\ell'$ are equal modulo 2; \textit{i.e.}, they have the same parity. 

\subsection*{E.2: Rotational symmetry}
Similarly, one can consider rotations around the $z$ axis by angle $\varphi = 2\pi/\mathfrak{n}_{\rm L}$. According to Eq.~\eqref{eq_translation_coeff}, the lattice sums can be written upon the rotation  as
\begin{align}
    \widehat{\Sigma}'_{\ell, m,\ell', m'}(\omega, \bm{0}) = \mathrm{e}^{\mathrm{i} 2\pi (m - m') / \mathfrak{n}_{\rm L}} \Sigma'_{\ell, m,\ell', m'}(\omega; \bm{0})\, .
\end{align}
One can clearly see that the lattice sum is preserved only if $(m - m') = s\mathfrak{n}$, where $s$ is an integer; that is, $m$ and $m'$ are equal modulo $\mathfrak{n}_{\rm L}$. 

\section{Calculation of the lattice field sums}\label{sec:sums}
Let us calculate lattice sums~\eqref{eq_sigma} in the far-field approximation $kr \gg 1$. First, we will use the integral expansion of the singular scalar spherical wave into plane waves~\cite{Wittmann1988Aug}, 
\begin{align}
\label{eq_sw_into_pw}
    \Psi^{(3)}_{\ell, m}(\omega; \mathbf{r}) = \frac{(-\mathrm{i})^{\ell}}{2 \pi } \iint\limits_{\mathbb{R}^2} \frac{\mathrm{d}k_x \mathrm{d}k_y}{k^2 \gamma} Y_{\ell, m}(\theta_{\mathbf{k}},\varphi_{\mathbf{k}})\mathrm{e}^{\mathrm{i}\mathbf{k}\cdot\mathbf{r}}\, ,
\end{align}
where $k = \omega/c$, $\gamma = \sqrt{1 - (k_x^2 + k_y^2) / k^2}$, and the unspecified component of the wave vector is $k_z = \pm k \gamma$ for $z \gtrless 0$.
Next, we utilize the Poisson summation formula 
\begin{align}
\label{eq_poisson}
    \sum_{\mathbf{R}} \mathrm{e}^{\mathrm{i}(\mathbf{k} - \mathbf{k}_{\parallel})\mathbf{R}} =  \frac{(2 \pi)^2}{A} \sum_{\mathbf{G}} \delta^{(2)}(\mathbf{k} - \mathbf{k}_{\parallel}- \mathbf{G})\, , 
\end{align}
where $A$ is the area of a unit cell, $\mathbf{G}$ is a vector of the reciprocal lattice, and $\delta^{(2)}$ is the 2D Dirac delta function, which eliminates the integration.  After combining Eqs.~\eqref{eq_sw_into_pw},~\eqref{eq_poisson}, and ~\eqref{eq_sigma}, we can express the lattice sum as
\begin{align}
\label{eq_sigma_lm}
    \Sigma_{\ell, m}(\omega, \mathbf{k}_{\parallel}; \mathbf{r}) = \frac{2 \pi}{k^2 A} (-\mathrm{i})^{\ell} \sum_{\mathbf{G}} Y_{\ell, m}(\theta_{\mathbf{k}},\varphi_{\mathbf{k}})\mathrm{e}^{\mathrm{i}\mathbf{k}\cdot\mathbf{r}}\, ,
\end{align}
where $\mathbf{k} = \mathbf{k}_{\parallel} + \mathbf{G} \pm \sqrt{k^2 - (\mathbf{k}_{\parallel} + \mathbf{G})^2}\widehat{\mathbf{z}}$ for $z \gtrless 0$.

If the wavelength in the background medium with $c_{\rm b}$ is longer than the lattice constant $L$, only waves with $\mathbf{G} = \bm{0}$ can propagate in the background medium. Hence, in the far-field approximation, Eqs.~\eqref{eq_sigma_lm} and~\eqref{eq_sigma_lm_subwave} coincide.

\section{Calculation of the real parts of lattice sums $\Sigma'_{\rm MM}$, $\Sigma'_{\rm MQ}$, and $\Sigma'_{\rm QQ}$}\label{sec:sums_real}
In the following, we calculate the real parts of the lattice sums that are present in Eq.~\eqref{eq_A}. Combining Eqs.~\eqref{eq_lattice_sums_Gamma},~\eqref{eq_translation_coeff}, and~\eqref{eq_sigma_lm_subwave}, one can obtain that
\begin{align}
\label{eq_ls_deriv_1}
\begin{aligned}
    & \Sigma_{\ell, 0, \ell, 0}'(\omega, \mathbf{0}) = \sqrt{4\pi } (2\ell+1)\sum\limits_{\substack{q = 0 \\ q \ \text{even}}}^{2 \ell} \mathrm{i}^{q} \sqrt{2q+1} \\ &\times \lim\limits_{\mathbf{r} \to \mathbf{0}}\left[ \Sigma_{q,0}(\omega, \mathbf{0}; \mathbf{r}) -  \Psi^{(3)}_{q,0}(\omega; \mathbf{r})\right] \begin{pmatrix}
            \ell& \ell & q \\
            0& 0& 0
        \end{pmatrix}^2\, .
\end{aligned}
\end{align}
The lattice field sums $\Sigma_{q,0}$ for a lattice in the $xy$ plane with a lattice constant $L < \lambda_{\rm b}$ read as
\begin{align}
\label{eq_ls_deriv_2}
    \lim\limits_{\mathbf{r} \to \mathbf{0}}\Sigma_{q, 0}(\omega, \mathbf{0}; \mathbf{r}) = (-\mathrm{i})^{q}\frac{2 \pi}{k^2 A}  \sqrt{\frac{2q + 1}{4\pi}} P_{\ell}^0(1)\, , 
\end{align}
where $P_{\ell}^0(1) \equiv 1$. For the second term in Eq.~\eqref{eq_ls_deriv_1}, we use Eq.~\eqref{eq_psi} and obtain that
\begin{align}
\label{eq_ls_deriv_3}
    \lim_{\mathbf{r} \to \mathbf{0}} \Re \left[ \Psi^{(3)}_{q,0}(\omega; \mathbf{r}) \right] = 
    \begin{cases}
        1/\sqrt{4\pi}, & q =0\, , \\
        0, & q\neq 0\, ,
    \end{cases}
\end{align}
while $\lim\limits_{\mathbf{r} \to \mathbf{0}} \Im \left[ \Psi^{(3)}_{q,0}(\omega; \mathbf{r}) \right]$ does not exist. Inserting Eqs.~\eqref{eq_ls_deriv_2} and~\eqref{eq_ls_deriv_3} into Eq.~\eqref{eq_ls_deriv_1} and using the properties of the Wigner's $3j$ symbols~\cite{Khersonskii1988}, we obtain that
\begin{align}
\begin{aligned}
    &\Re\left [\Sigma_{\ell, 0, \ell, 0}'(\omega, \mathbf{0})\right] = \\&-1 + (2\ell + 1) \frac{\lambda_{\rm b}^2}{2 \pi A} \underbrace{ \sum\limits_{\substack{q = 0 \\ q \ \text{even}}}^{2 \ell}(2q + 1)\begin{pmatrix}
            \ell& \ell & q \\
            0& 0& 0
        \end{pmatrix}^2}_{=1}\, ,
\end{aligned}
\end{align}
where $\lambda_{\rm b} =2\pi c_{\rm b} / \omega$ and $A = L^2$ for the square lattice.
Similarly, we can calculate the real part of the lattice sums with even difference $(\ell' - \ell)$ as
\begin{align}
\label{eq_ls_deriv_4}
\begin{aligned}
    \Re\left [ \Sigma_{\ell, 0, \ell', 0}'(\omega, \mathbf{0}) \right] = \sqrt{(2\ell+1)(2\ell'+1)} \mathrm{i}^{\ell'-\ell}\frac{\lambda_{\rm b}^2}{2 \pi A} \\ \times \underbrace{\sum\limits_{\substack{q = \ell'-\ell \\ q \ \text{even}}}^{\ell + \ell'}  (2q+1)   \begin{pmatrix}
            \ell& \ell' & q \\
            0& 0& 0
        \end{pmatrix}^2}_{=1}\, .
\end{aligned}
\end{align}
Hence,
\begin{align}
\label{eq_re_ls}
\begin{aligned}
    \Re\left [ \Sigma_{\rm MM}'(\omega, \mathbf{0}) \right] &\equiv \Re\left [ \Sigma_{0, 0, 0, 0}'(\omega, \mathbf{0}) \right] = -1 + \frac{\lambda_{\rm b}^2}{2 \pi A}\, , \\
    \Re\left [ \Sigma_{\rm QQ}'(\omega, \mathbf{0}) \right] &\equiv \Re\left [ \Sigma_{2, 0, 2, 0}'(\omega, \mathbf{0}) \right] = -1 + \frac{5\lambda_{\rm b}^2}{2 \pi A}\, , \\
    \Re\left [ \Sigma_{\rm MQ}'(\omega, \mathbf{0}) \right] &\equiv \Re\left [ \Sigma_{0, 0, 2, 0}'(\omega, \mathbf{0}) \right] = -\frac{\sqrt{5}\lambda_{\rm b}^2}{2 \pi A}\, .
\end{aligned}
\end{align}

\section{Relationship between the BIC conditions}\label{sec:conditions}
In this section, we show the equivalence of BIC conditions~\eqref{eq_det} and~\eqref{eq_criterion1} for matrix~\eqref{eq_A}. First, we introduce the monopole and quadrupole T-matrix elements of a sphere modified due to the monopole-monopole and quadrupole-quadrupole coupling, respectively,
\begin{align}
\begin{aligned}
\label{eq_Tmod}
      \frac{1}{T_{\rm MM}^{\rm mod}} &= \frac{1}{T_{\rm MM}} -  \Sigma'_{\rm MM}\, , \\
      \frac{1}{T_{\rm QQ}^{\rm mod}} &= \frac{1}{T_{\rm QQ}} -  \Sigma'_{\rm QQ}\, .
\end{aligned}      
\end{align}
Considering Eqs.~\eqref{eq_re_inv_mie} and~\eqref{eq_re_ls}, one can see that for $L < \lambda_{\rm b}$, the real parts of Eq.~\eqref{eq_Tmod} are related to each other as
\begin{align}
\label{eq_re_equiv_1}
    \Re\left[ \frac{1}{T_{\rm MM}^{\rm mod}}\right] \equiv 5 \Re\left[ \frac{1}{T_{\rm QQ}^{\rm mod}}\right].
\end{align}
The determinant of Eq.~\eqref{eq_A} being zero implies that
\begin{align}
\label{eq_detA_analytical}
  \frac{1}{T_{\rm MM}^{\rm mod}} \times \frac{1}{T_{\rm QQ}^{\rm mod}} = \Sigma'_{\rm MQ}  \times \Sigma'_{\rm MQ}\, .
\end{align}
Due to Eq.~\eqref{eq_re_ls}, we also have for $L < \lambda_{\rm b}$ that
\begin{align}
\label{eq_re_equiv_2}
    \Re\left[ \frac{1}{T_{\rm MM}^{\rm mod}}\right] \equiv \sqrt{5} \Re \left[\Sigma'_{\rm MQ}\right]\, .
\end{align}
Hence, the solution to Eq.~\eqref{eq_detA_analytical} reads as
\begin{align}
\label{eq_detA_solution}
    \frac{1}{T_{\rm MM}^{\rm mod}} = \frac{1}{\sqrt{5}} \Sigma'_{\rm MQ}, \ \text{and} \
    \frac{1}{T_{\rm QQ}^{\rm mod}} = \sqrt{5} \Sigma'_{\rm MQ}\, .
\end{align}
Indeed, considering~\eqref{eq_re_equiv_1} and~\eqref{eq_re_equiv_2}, we can write that $1/T_{\rm MM}^{\rm mod} = A + \mathrm{i}B$, $1/T_{\rm QQ}^{\rm mod} = 5A + \mathrm{i}C$, and $\Sigma'_{\rm MQ} = \sqrt{5} A + \mathrm{i}D$, where $A$, $B$, $C$, and $D$ are real-valued numbers, and plug them into Eq.~\eqref{eq_detA_analytical}. Equating the real and imaginary parts of the left and right sides of the equation yields the following system of equations:
\begin{align}
    \begin{cases}
        5B + C = 2\sqrt{5}D\, ,\\
        BC =D^2\, .
    \end{cases}
\end{align}
The solution is $B = D/\sqrt{5}$ and $C = D\sqrt{5}$, and we obtain Eq.~\eqref{eq_detA_solution}.
If an eigenmode condition~\eqref{eq_detA_solution} is satisfied for $\omega \in \mathbb{R}$, then the lattice mode is a BIC.

If we rewrite Eqs.~\eqref{eq_lattice_modes} using~\eqref{eq_A} and~\eqref{eq_Tmod}, we obtain that
\begin{align}
    \begin{cases}
        1/T_{\rm MM}^{\rm mod} a^{\rm eff}_{\rm M} - \Sigma'_{\rm MQ} a^{\rm eff}_{\rm Q} = 0\, , \\
        - \Sigma'_{\rm MQ} a^{\rm eff}_{\rm M} +1/T_{\rm QQ}^{\rm mod} a^{\rm eff}_{\rm Q} = 0\, .
    \end{cases}
\end{align}
Under BIC condition~\eqref{eq_detA_solution}, we immediately obtain Eq.~\eqref{eq_criterion1}, \textit{i.e.}, $\left( a^{\rm eff}_{\rm M} - \sqrt{5}a^{\rm eff}_{\rm Q}\right) = 0$ corresponding to the far-field intensity $\mathcal{P}_{\pm} = 0$.
Note that, for a normally incident plane wave of amplitude $p_0$, Eq.~\eqref{eq_mse_lattice_2} under the BIC condition gives that $\left( a^{\rm eff}_{\rm M} - \sqrt{5}a^{\rm eff}_{\rm Q}\right) \propto p_0$ and consequently $\mathcal{P}_{\pm} \neq 0$. Hence, the BIC cannot be excited by a normally incident plane wave.

\begin{figure}
    \centering
    \includegraphics[scale=0.4]{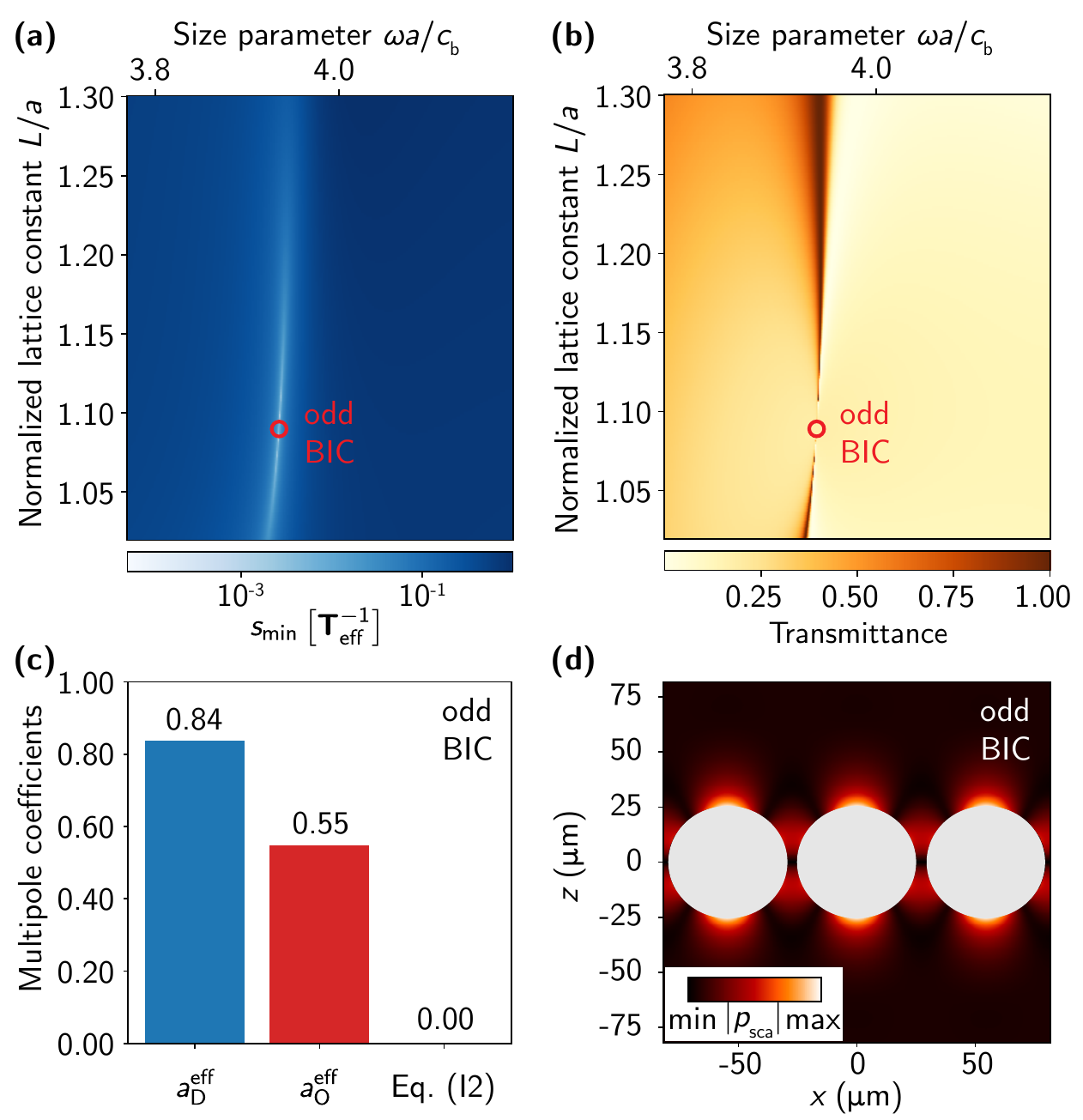}
    \caption{Odd BIC in the metasurface of spherical resonators depicted in Fig.~\ref{fig:sketch}(a). (a) The lowest singular value of the inverse effective T-matrix in the dipole-octupole approximation~\eqref{eq_A_odd}.  (b) Transmittance of the lattice $|t|^2$ for a normally incident pressure plane wave with $\ell_{\rm max} = 6$ in Eq.~\eqref{eq_t}. The quantities in panels (a) and (b) are plotted as a function of the normalized lattice constant $L/a$ and size parameter $\omega a / c_{\rm b}$, where $a = 50$ $\mu$m and $c_{\rm b} = 1500$ m/s (the other parameters are listed in the text). The red circles indicate the lattice constant and the frequency, for which the odd BIC emerges. (c) Effective zonal dipole and octupole moments of the odd BIC. (d) Normalized absolute value of the pressure field generated by the BIC in the $xz$ plane outside the spheres.}
    \label{fig:bic_odd}
\end{figure}

\section{Odd $\Gamma$-point BIC in the dipole-octupole approximation}\label{sec:odd}
In the main text, we consider the BIC formed by the zonal multipoles of even parity. Similarly, we can consider the BIC formed by the multipoles with odd parity, the dipole and octupole. The frequency and lattice constant of the odd-parity BIC can also be determined by Eq.~\eqref{eq_smin} with the following inverse effective T-matrix:
\begin{align}
\label{eq_A_odd}
    \mathbf{T}^{-1}_{\rm eff}(\omega) = 
    \begin{pmatrix}
        T_{\rm DD}^{-1} -  \Sigma'_{\rm DD} & -\Sigma'_{\rm DO} \\
        -\Sigma'_{\rm DO} & T_{\rm OO}^{-1}  - \Sigma'_{\rm OO}
    \end{pmatrix}\, ,
\end{align}
where, for a sphere, $T_{\rm DD} \equiv \mathfrak{a}_1$ and $T_{\rm OO} \equiv \mathfrak{a}_3$ [see Eq.~\eqref{eq_mie_coeffs}]. Here, the lattice sums are $\Sigma'_{\rm DD} \equiv \Sigma'_{\rm 1,0,1,0}(\omega,\bm{0}), \Sigma'_{\rm DO} \equiv \Sigma'_{\rm 0,0,3,0}(\omega,\bm{0}),$ and $\Sigma'_{\rm OO} \equiv \Sigma'_{\rm 3,0,3,0}(\omega,\bm{0})$ [see Eq.~\eqref{eq_lattice_sums}]. Note that additional off-diagonal terms due to $T_{\rm DO} = T_{\rm OD} \neq 0$ appear in Eq.~\eqref{eq_A_odd} for nonspherical particles. Using the same approach as in Appendix~\ref{sec:conditions}, we can derive Eq.~\eqref{eq_criterion_odd} from condition $\left[ \det \mathbf{T}^{-1}_{\rm eff}(\omega) = 0 \right]$ with $\omega \in \mathbb{R}$. Figure~\ref{fig:bic_odd} presents the same content as Fig.~\ref{fig:lattice_spheres}, but for the odd BIC. 

\section{T-matrix approach to finite-size arrays of resonators}\label{sec:Tmatrix_finite}
The T-matrix-based formalism described in Sec.~\ref{sec:theory} for infinite lattices can be adopted to finite-size arrangements of $N_{\rm tot}$ scatterers. In the latter case, the total scattered field reads as
\begin{align}
    p_{\rm sca}(\omega; \mathbf{r}) &= \sum_{i = 1}^{N_{\rm tot}} \sum_{\ell,m}  a_{i,\ell, m} \Psi^{(3)}_{\ell, m}(\omega; \mathbf{r} - \mathbf{r}_i)\, ,
\end{align}
where $a_{i,\ell, m}$ are unknown coefficients. To calculate them, we expand a given incident field around positions $\mathbf{r}_i$ of the scatterers,
\begin{align}
    p_{\rm inc}(\omega; \mathbf{r}) &= \sum_{\ell,m}  b_{i,\ell, m}(\omega) \Psi^{(1)}_{\ell, m}(\omega; \mathbf{r} - \mathbf{r}_i)\, .
\end{align}
Next, we can write the local field coefficients as [cf. Eq.~\eqref{eq_coeffs_local}]
\begin{align}
\begin{aligned}
    \widetilde{b}_{i,\ell, m}(\omega) = &b_{i,\ell, m}(\omega) \\ &+ \sum_{\ell',m'} \sum_{j \neq i} \mathcal{C}^{(3)}_{\ell, m,\ell', m'}(\omega; \mathbf{r}_i -\mathbf{r}_j) a_{j,\ell', m'} (\omega)
\end{aligned}
\end{align}
and substitute them into Eq.~\eqref{eq_Tmatrix} for each scatterer to obtain the following system of linear equations:
\begin{align}
\label{eq_mse}
    \mathbf{a}_i = \mathbf{T}_i \left[\mathbf{b}_i + \sum_{j \neq i} \bm{\mathcal{C}}^{(3)}_{ij} \mathbf{a}_j\right]\, ,
\end{align}
where $\mathbf{a}_i \equiv \mathbf{a}_i(\omega)$, $\mathbf{b}_i \equiv \mathbf{b}_i(\omega)$, $\mathbf{T}_i \equiv \mathbf{T}_i(\omega)$ is the acoustic T-matrix of the $i$th scatterer, and $\bm{\mathcal{C}}^{(3)}_{ij}$ is the matrix that contains the translation coefficients $\mathcal{C}^{(3)}_{\ell, m,\ell', m'}(\omega; \mathbf{r}_i -\mathbf{r}_j)$.

\section{Description of the FEM model}\label{sec:fem}
Simulations are performed using the finite-element method (FEM) with the \textit{pressure acoustics, frequency domain} interface in COMSOL Multiphysics\texttrademark \ (version 5.5). The computational domain consists of a rigid-walled parallelepiped with dimensions $120 \times 60 \times 60$ $\mu$m, within which a spherical object of radius 25 $\mu$m is centered. The surrounding water fills the region between the surface of the sphere and the inner boundaries of the parallelepiped. To reduce the computational cost, symmetry boundary conditions are applied on selected faces of the parallelepiped. Port boundary conditions are used to excite acoustic waves entering the structure. The rectangular port option is selected, and the first four $(m, n)$ modes of a waveguide with a rectangular cross-section are included. The computational domain is discretized using tetrahedral finite elements with an \textit{extra fine} element size.

\bibliography{main}
\end{document}